%% file: main.tex
\documentclass[aps,physrev,preprint,groupedaddress]{revtex4-2}
\usepackage{graphicx}
\usepackage{epstopdf,epsfig}
\usepackage[]{subcaption} 
\usepackage{multirow}
\usepackage{array}
\newcolumntype{x}[1]{>{\centering\arraybackslash\hspace{0pt}}p{#1}}
\usepackage{amsmath}
\usepackage{enumitem} 
\usepackage{placeins}
\usepackage{newtxtext}
\usepackage{newtxmath}
\usepackage{natbib}
\usepackage{hyperref}
\hypersetup{
    colorlinks = true,
    linkcolor  = blue,
    urlcolor   = blue,
    citecolor  = blue,
}
\usepackage{todonotes}

\newcommand{\RomanNumeralCaps}[1]
\linenumbers

\begin{document}

\title{Time-varying wind-turbine wakes at high Reynolds numbers}

\author{Nathaniel J. Wei}
  \email{Email for correspondence: njwei@seas.upenn.edu}
  \affiliation{Mechanical Engineering and Applied Mechanics, University of Pennsylvania, Philadelphia, PA 19104, USA}
  \affiliation{Andlinger Center for Energy and the Environment, Princeton University, Princeton, NJ 08544, USA}

\author{Adina Y. Fleisher}
\affiliation{Mechanical and Aerospace Engineering, Princeton University, Princeton, NJ 08544, USA}

\author{John W. Kurelek}
\affiliation{Mechanical and Materials Engineering, Queen's University, Kingston, ON, Canada K7L 3N6}

\author{Marcus N. Hultmark}
\affiliation{Mechanical and Aerospace Engineering, Princeton University, Princeton, NJ 08544, USA}

\date{\today}

\begin{abstract}
Wind turbines operating in the atmospheric boundary layer are constantly exposed to time-varying flow conditions. These disturbances often occur on similar time scales to wind-turbine controllers, which may interfere with wind-farm control strategies that operate under steady-flow assumptions. This study aims to investigate the significance of such time variations on wind-turbine wake dynamics, focusing on slow time scales representative of quasi-steady processes in large wind farms. Experiments are conducted at near utility-scale Reynolds numbers ($Re_D=4\times10^6$) in a pressurized-air wind tunnel, with a wind turbine forced in periodic rotation-rate oscillations by means of a time-varying generator torque at low Strouhal numbers ($St=0.04$). Flow measurements in the wake of the turbine demonstrate that disturbances propagate through the wake as traveling waves, which are advected nonlinearly at the velocity of the wake rather than that of the free stream. The wake behavior can be described in a quasi-steady manner, but only after wake advection is accounted for by a Lagrangian transformation. Even in the quasi-steady regime, the spatiotemporal evolution of the wake can be controlled by independently varying the turbine thrust and tip-speed ratio. The results suggest that wake advection is important to consider for wind-farm modeling and control, and that time-varying control may allow wind-turbine wake interactions to be tuned even in nominally quasi-steady conditions for optimal wind-farm performance.
\end{abstract}

\maketitle


\input{s1_introduction}
 
\input{s2_theory}

\input{s3_methods}

\input{s4_results}

\input{s5_conclusions}

\section*{Acknowledgements}{The authors would like to thank Dan Hoffman for his tireless assistance with the experimental apparatus and facilities.}

\section*{Funding}{N.J.W.\ was supported by a Distinguished Postdoctoral Fellowship from the Andlinger Center for Energy and the Environment at Princeton University. J.W.K.\ gratefully acknowledges funding support from the Natural Sciences and Engineering Research Council of Canada.}

\section*{Declaration of interests}{The authors report no conflict of interest.}

\section*{Data availability statement}{The data that support the findings of this study are available upon reasonable request.}

\section*{Author ORCID}{N.\ J.\ Wei, \href{https://orcid.org/0000-0001-5846-6485}{https://orcid.org/0000-0001-5846-6485}; A.\ Y.\ Fleisher, \href{https://orcid.org/0009-0001-3034-2904}{https://orcid.org/0009-0001-3034-2904}; J.\ W.\ Kurelek, \href{https://orcid.org/0000-0002-8889-3119}{https://orcid.org/0000-0002-8889-3119}; M.\ N.\ Hultmark, \href{https://orcid.org/0000-0002-5070-3711}{https://orcid.org/0000-0002-5070-3711}.}

\section*{Author contributions}{N.\ J.\ Wei: Conceptualization, theory, methodology, investigation, analysis, writing -- original draft. A.\ Y.\ Fleisher: Conceptualization, methodology, investigation, analysis, writing -- review \& editing. J.\ W.\ Kurelek: Conceptualization, methodology, writing -- review \& editing. M.\ N.\ Hultmark: Conceptualization, supervision, writing -- review \& editing.}

\input{s9_appendices}

\bibliographystyle{unsrtnat}
\bibliography{DIC.bib}

\end{document}

%% file: s1_introduction.tex
\section{Introduction}
\label{sec:intro}

The rapid expansion of wind energy is a critical task for the global transition towards a more environmentally sustainable energy economy. A major challenge facing large-scale wind energy development is the downstream impacts of wind-turbine and wind-farm wakes. In large wind farms, power-generation losses due to wakes from upstream turbines are often on the order of 10 to 20\% \citep{barthelmie_modelling_2009}, with individual turbine losses approaching 30 to 50\% in some cases \citep{barthelmie_quantifying_2010,howland_wind_2019,pryor_wind_2021}. As wind farms grow larger and more prevalent worldwide, these wake losses represent inefficiencies that, if reduced, could result in significant improvements in power generation and reductions in the levelized cost of energy for wind farms. The study and control of wind-turbine wake dynamics are therefore key to advancing strategies to mitigate the long-term effects of anthropogenic climate change.

As wind turbines and wind farms continue to grow in size, the evolution of wind-turbine wakes span increasingly large areas. Mechanical turbulence in the atmospheric boundary layer typically has a spectral peak at time scales on the order of 1 to 10 minutes \citep{van_der_hoven_power_1957,stull_introduction_1988}. If the transient disturbances represented by this peak propagate at approximately the local wind speed, it may take several minutes for these phenomena to travel from an upstream row of turbines to a downstream one, thus creating significant advective time lags between turbines within the array \citep{brugger_field_2022,foloppe_development_2022}. Since wind-farm controllers are often designed using quasi-steady assumptions and operate on the order of minutes \citep{starke_dynamic_2024}, these time-varying wake dynamics may interfere with control and optimization strategies, potentially leading to power-generation losses and increased unsteady fatigue loads on critical structural components. Additional time-varying forces introduced by very large-scale motions (VLSMs) in the atmospheric boundary layer and the passage of mesoscale weather systems such as fronts, squall lines, and convective complexes further complicate the ever-shifting flows within large wind farms \citep{veers_grand_2019,shaw_scientific_2022,kosovic_impact_2026}.

In addition to the effects of wake advection, several other time-varying dynamics are present in wind-turbine wakes. Even in steady, uniform inflow, instability of the wake can lead to dynamic wake meandering, which manifests as lateral perturbations in the wake as it evolves downstream \citep{yang_review_2019}. When changes in inflow conditions or turbine operation occur, additional dynamic wake modulation can be excited in response \citep{abraham_dynamic_2020,abraham_mechanisms_2021}. For floating offshore wind turbines, platform oscillations can further affect the unsteady evolution of the wake, in some cases generating coherent structures that accelerate wake recovery \citep{rockel_wake_2016,van_den_berg_dynamic_2023,messmer_enhanced_2024,wei_wake_2024,fontanella_wake_2025,messmer_role_2025}. Recent developments in dynamic wake control have sought to exploit some of these unsteady mechanisms in wind-turbine wakes to accelerate wake recovery and thereby enhance wind-farm power density \citep[\textit{cf}.,][]{meyers_wind_2022}. For example, \citet{brown_accelerated_2022} demonstrated intracycle rotation-rate perturbations can encourage mutual-inductance instabilities that break down the helical tip vortices shed by the turbine blades, thus shortening the near wake. In a different approach, known as dynamic induction control, the thrust force of the turbine is perturbed in periodic oscillations \citep{goit_optimal_2015,munters_optimal_2017,munters_towards_2018}. This can be done by periodically varying the collective pitch angle of the turbine blades \citep[\textit{e.g.},][]{frederik_periodic_2020,van_der_hoek_experimental_2022}, varying the individual blade pitch azimuthally to create a nonuniform rotating thrust profile \citep[\textit{e.g.},][]{frederik_helix_2020,vander_hoek_maximizing_2024}, or moving the turbine itself in periodic oscillations, which is a possibility for floating offshore wind turbines \citep{messmer_enhanced_2024,wei_wake_2024}. Unsteady dynamics may also decouple the turbine thrust from the tip-speed ratio due to the effects of inertia, generator loading, and unsteady blade aerodynamics \citep{el_makdah_scaling_2021,wei_phase-averaged_2022,wei_power-generation_2023}.


Modeling these time-varying dynamics for wind-farm simulation and control remains an active area of research. Many existing engineering models for wind-farm design and layout optimization rely on quasi-steady assumptions, such as FLORIS \citep{gebraad_control-oriented_2014,gebraad_wind_2016} and FAST.Farm \citep{jonkman_development_2017}. Recent extensions have been made to these types of engineering wake models to include the effects of wake advection \citep{jonkman_fastfarm_2021,becker_revised_2022}. Such dynamic corrections typically integrate the trajectories of individual wake elements in a Lagrangian approach, using either the mean wind speed or the local wake velocity \citep{kheirabadi_low-fidelity_2021,foloppe_development_2022,song_novel_2025}. This can be done for both streamwise and lateral wake evolution, with the latter enabling modeling under yaw misalignment \citep{deng_yaw_2023,micheletto_experimental_2023}. Additional models have been derived to capture unsteady wake dynamics beyond pure advection, including dynamic wake meandering models \citep{larsen_wake_2008,madsen_calibration_2010}, a one-dimensional model for nonlinear wake advection and deformations under unsteady turbine forcing \citep{wei_wake_2024}, and resolvent-based modal approaches that connect unsteady turbine dynamics to the time evolution and recovery of the wake \citep{li_resolvent-based_2024,li_self-consistent_2025}.

While these recently developed modeling approaches parameterize time-varying wake dynamics, scant experimental data at field-relevant Reynolds numbers exist to evaluate their predictions. Most lab-scale experiments on wake dynamics have been conducted at diameter-based Reynolds numbers of $Re_D = \mathcal{O}(10^5)$, which lie one to two orders of magnitude below those of utility-scale wind turbines \citep[\textit{e.g.},][]{krogstad_performance_2012,zhang_near-wake_2012,bartl_wind_2018}, and thus rely on Reynolds-number invariance arguments that remain to be proven definitively \citep{chamorro_reynolds_2012,miller_horizontal_2019}. Field measurements of the wakes of utility-scale wind turbines have been conducted using meteorological towers \citep[\textit{e.g.},][]{duckworth_investigation_2008}, LiDAR \citep[\textit{e.g.},][]{iungo_field_2013,iungo_volumetric_2014,aitken_quantifying_2014,heisel_spectral_2018}, and even natural snowfall \citep{hong_natural_2014,dasari_near-wake_2019,abraham_dynamic_2020}. Most of these studies have been limited in spatial or temporal resolution, and could not isolate the cause of any observed time-varying wake dynamics due to lack of control over inflow conditions and turbine operation. Experiments conducted in pressurized-air wind tunnels can fill these knowledge gaps, as such studies are capable of achieving dynamic similarity to field conditions while maintaining full control over flow properties and turbine operation, while offering the high levels of spatial and temporal resolution typical of laboratory experiments \citep{miller_horizontal_2019,pique_dominant_2022,pique_laboratory_2022,pique_understanding_2025}.

Accordingly, the present study investigates the effects of time-varying forcing on wind-turbine wakes at near utility-scale Reynolds numbers, at time scales relevant to large wind farms, via flow measurements in a pressurized-air wind tunnel. The work seeks to evaluate two primary hypotheses: first, that wake advection is a critical factor for wake modeling and wind-farm control, and second, that the operation of the turbine can be leveraged to control the structure and temporal evolution of the wake. These two hypotheses are tested using relatively slow forcing time scales that would be considered quasi-steady in typical wind-farm operation, in order to demonstrate that time-varying dynamics are important even in scenarios that are typically assumed to be steady.

The work is structured as follows. First, in \S\ref{sec:theory}, theoretical considerations are employed to demonstrate the importance of wake advection for this problem. The experimental apparatus and method for forcing the turbine and wake are described in \S\ref{sec:methods}. Results from the flow measurements, documented in \S\ref{sec:results}, highlight the effects of advection on the wake structure and evolution. The experiments also demonstrate how control of the turbine thrust and tip-speed ratio can alter the evolution of the wake, even at slow forcing time scales. Implications of these results for wind-farm modeling and control are discussed in \S\ref{sec:conclusions}.

%% file: s2_theory.tex
\section{Theoretical considerations}\label{sec:theory}


\subsection{Wind-turbine wake dynamics in steady flow}

A large body of research on wind-turbine wake dynamics in steady flow exists in the literature \citep[\textit{cf.},][]{stevens_flow_2017,porte-agel_wind-turbine_2020}. According to one-dimensional (1D) momentum theory, the velocity deficit in the wake of a turbine is a function of the thrust coefficient, defined as

\begin{equation}
    C_t = \frac{F_x}{\frac{1}{2}\rho U_\infty^2 A},
    \label{eqn:Ct}
\end{equation}

\noindent where $F_x$ is the streamwise force on the turbine rotor, $\rho$ is the fluid density, $U_\infty$ is the free-stream velocity, $A=\frac{\pi}{4}D^2$ is the swept area of the turbine, and $D$ is the turbine diameter. Conservation relations dictate that, at least for relatively low axial induction factors, a higher thrust coefficient leads to a stronger wake-velocity deficit. A typical axisymmetric turbulent wake will recover from this initial velocity deficit as the flow evolves downstream, driven by the turbulent entrainment of momentum from the undisturbed free stream surrounding the wake \citep{uberoi_turbulent_1970}. However, unlike the canonical bluff-body case, wind-turbine wakes feature a system of helical tip vortices that plays a key role in defining the near, intermediate, and far wake regions. The helical tip vortices are shed by the rotating turbine blades and form a cylindrical `shell' that shields the near wake from the free stream. These vortex elements undergo a mutual-inductance instability in which neighboring vortices pair up and eventually break down \citep{widnall_stability_1972,okulov_stability_2007,sarmast_mutual_2014,sorensen_simulation_2015,lignarolo_tip-vortex_2015}. These dynamics are dependent on the tip-speed ratio of the turbine, defined as

\begin{equation}
    \lambda = \frac{\Omega D}{2U_\infty},
    \label{eqn:TSR}
\end{equation}

\noindent where $\Omega$ is the rotation rate of the turbine. The breakdown of the helical tip vortex, which typically occurs at a downstream distance of $2\lesssim x/D \lesssim 4$, marks the transition into the intermediate wake, in which the shear layers grow inwards from the wake boundary toward the centerline. The far-wake region begins after the shear layers reach the centerline and all signatures of the near-wake vortex system, and thus properties of the rotor, can no longer be identified \citep{vermeer_wind_2003}. Therefore, modeling approaches that fail to capture turbine rotation and the helical tip vortex system in the near wake, such as actuator-disc models, can only capture far-wake behaviors \citep{aubrun_wind_2013,lignarolo_tip-vortex_2015,kurelek_performance_2023}. In steady flow, the far wake behaves in a self-similar manner \citep[\textit{e.g.},][]{bastankhah_new_2014}. Characteristics of the tip-vortex system can be related to the turbine thrust coefficient, for instance by Biot-Savart induction \citep[\textit{e.g.},][]{segalini_simplified_2013,de_vaal_validation_2014} or by impulse theory \citep{limacher_relationship_2022}. This schematic overview of steady-flow wake dynamics suggests that a full account of wind-turbine wake recovery must consider the effects of both thrust and tip-speed ratio.

\subsection{The effects of time-varying forcing on wind-turbine wakes}
\label{sec:theory_waves}

Time-varying forcing, due to changes either in the inflow or in turbine operation, alter the steady-flow behavior of the wake. To demonstrate this, we write the equation for conservation of momentum in the streamwise direction for an axisymmetric wake, assuming no swirl so that the nontrivial velocity components are in the axial ($u_x$) and radial ($u_r$) directions.

\begin{equation}
    \frac{\partial u_x}{\partial t} + u_x \frac{\partial u_x}{\partial x} + u_r \frac{\partial u_x}{\partial r} = -\frac{1}{\rho} \frac{\partial p}{\partial x} + \nu\left(\frac{1}{r} \frac{\partial}{\partial r}\left(r \frac{\partial u_x}{\partial r}\right) + \frac{\partial^2 u_x}{\partial x^2}\right)
    \label{eqn:momentum}
\end{equation}

We perform a triple decomposition of the flow field $u_i$ into a time-averaged component $\overline{U}_i$, a harmonic time-varying perturbation $\tilde{u}_i$, and turbulent fluctuations $u_i'$. The phase-averaged velocity is defined as $U_i = \overline{U}_i + \tilde{u}_i$. Phase-averaging Eq.\ \ref{eqn:momentum} yields

\begin{multline}
    \frac{\partial U_x}{\partial t} + U_x \frac{\partial U_x}{\partial x} + U_r \frac{\partial U_x}{\partial r} = -\frac{1}{\rho} \frac{\partial P}{\partial x} + \nu\left(\frac{1}{r} \frac{\partial}{\partial r}\left(r \frac{\partial U_x}{\partial r}\right) + \frac{\partial^2 U_x}{\partial x^2}\right) \\
    - \frac{\partial}{\partial x} \langle u_x' u_x' \rangle - \frac{1}{r}\frac{\partial}{\partial r} \left( r \langle u_x' u_r' \rangle \right),
    \label{eqn:momentum_phaseavg}
\end{multline}

\noindent where angle brackets denote phase-averaged statistics. Following \citet{iungo_parabolic_2018} and \citet{wei_wake_2024}, we employ Boussinesq hypotheses for the turbulence statistics:

\begin{align}
    -\langle u_x' u_x' \rangle &= 2\nu_T \frac{\partial U_x}{\partial x} - \frac{2}{3}k; \label{eqn:uxux} \\
    -\langle u_x' u_r' \rangle &= \nu_T \left(\frac{\partial U_x}{\partial r} + \frac{\partial U_r}{\partial x}\right), \label{eqn:uxur}
\end{align}

\noindent where $k$ is the turbulence kinetic energy. Assuming that the eddy viscosity $\nu_T$ is not a function of space, we arrive at the phase-averaged momentum equation,

\begin{multline}
    \frac{\partial U_x}{\partial t} + U_x \frac{\partial U_x}{\partial x} + U_r \frac{\partial U_x}{\partial r} = -\frac{1}{\rho} \frac{\partial P}{\partial x} + \nu\left(\frac{1}{r} \frac{\partial}{\partial r}\left(r \frac{\partial U_x}{\partial r}\right) + \frac{\partial^2 U_x}{\partial x^2}\right) \\
    + \nu_T \left( 2\frac{\partial^2 U_x}{\partial x^2} + \frac{\partial^2 U_x}{\partial r^2} + \frac{\partial^2 U_r}{\partial r \partial x} + \frac{1}{r}\frac{\partial U_x}{\partial r} + \frac{1}{r}\frac{\partial U_r}{\partial x} \right) - \frac{2}{3} \frac{\partial k}{\partial x}.
    \label{eqn:momentum_RANS}
\end{multline}

To highlight the effects of time variations on wind-turbine wakes, we now make a series of simplifying assumptions. First, since pressure recovers rapidly in the near wake, we may neglect the pressure-gradient term in the wake at downstream distances greater than 1 to 2 turbine diameters \citep{bempedelis_analytical_2022}. This removes the nonlocal term and parabolizes the governing equations. Secondly, to focus on streamwise dynamics, we assume radial homogeneity and quasi-parallel flow, removing terms involving $\partial/\partial r$ and $U_r$. These assumptions are justified by the fact that radial velocities in the wake are generally smaller than streamwise velocities, and the length scales of spatial variations in $r$ are smaller than those in $x$. Under these quasi-1D assumptions, Eq.\ \ref{eqn:uxux} reduces to

\begin{equation}
    \langle u_x' u_x' \rangle \approx \nu_T \frac{\partial U_x}{\partial x},
    \label{eqn:uxux_1D}
\end{equation}

\noindent further simplifying Eq.\ \ref{eqn:momentum_RANS}. These are undoubtedly strong oversimplifications for turbine wakes, but for this work they serve to highlight the effects of time variations on the streamwise wake evolution. Assuming that the turbulent viscosity dominates over viscous diffusion, we have

\begin{equation}
    \frac{\partial U_x}{\partial t} + U_x \frac{\partial U_x}{\partial x} + \nu_T \frac{\partial^2 U_x}{\partial x^2} = 0.
    \label{eqn:burgers}
\end{equation}

\noindent As \citet{wei_wake_2024} have shown using a similar derivation, this simplified model is a viscous Burger's equation: a nonlinear parabolic partial differential equation that admits traveling waves as solutions. Their towing-tank experiments demonstrated that this reduced-order model captures the nonlinear growth and steepening of waves in the wake radius, as a function of large-amplitude oscillations in turbine thrust. By contrast, for small-amplitude oscillations ($\tilde{u}_x << \overline{U}_x$), Eq.\ \ref{eqn:burgers} further simplifies to

\begin{equation}
    \frac{\partial \tilde{u}_x}{\partial t} + \overline{U}_x \frac{\partial \tilde{u}_x}{\partial x} + \tilde{u}_x \frac{\partial \overline{U}_x}{\partial x} + \nu_T \frac{\partial^2 \tilde{u}_x}{\partial x^2} = 0,
    \label{eqn:burgers_tilde}
\end{equation}

\noindent which is an advection-diffusion equation for the velocity perturbation $\tilde{u}_x$. This form shows that these perturbations will be advected through the wake at a speed dictated by the local wake velocity at any given streamwise location. This suggests that wake perturbations will accelerate as the wake recovers further downstream.

The above analysis demonstrates the importance of advection for time-varying wind-turbine wakes, even at slow forcing time scales. It also underscores the nature of the propagation of disturbances into the wake, implying that changes in thrust or tip-speed ratio will advect downstream through the intermediate and far wake at a speed governed by the local wake velocity.

%% file: s3_methods.tex
\section{Experimental methods}\label{sec:methods}

\subsection{Experimental apparatus}\label{sec:methods_apparatus}

\begin{figure}
\centering
  \includegraphics[width=0.5\textwidth]{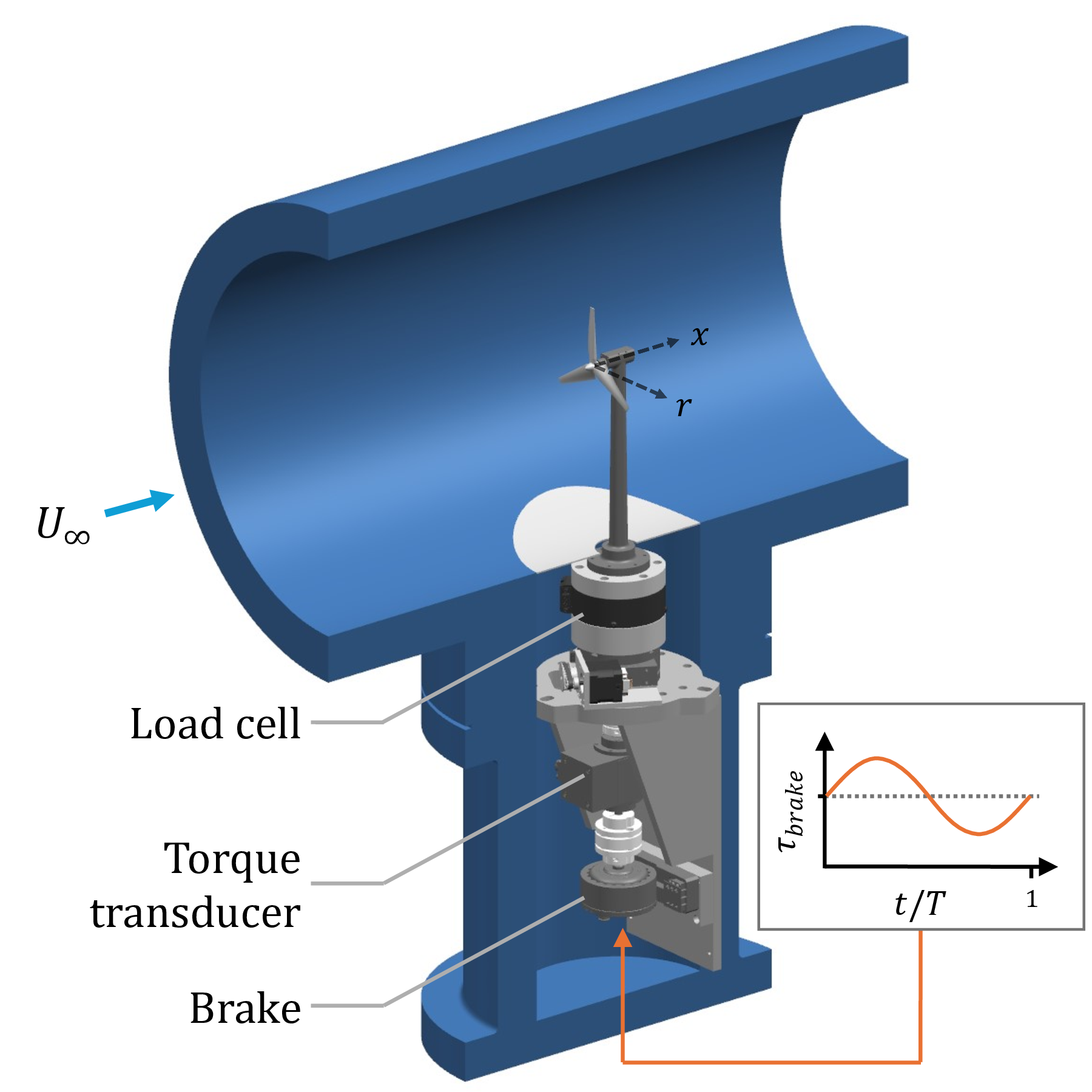}
  \caption{Sectional view of the HRTF test section, including turbine model ($D=15$ cm) and measurement stack. A sample time-varying torque waveform supplied to the brake is shown in an inset.}
\label{fig:setup}
\end{figure}

Experiments were conducted in the High Reynolds number Test Facility (HRTF) at Princeton University, which uses pressurized air as the working fluid to obtain Reynolds numbers comparable to those of utility-scale wind turbines while maintaining realistic tip-speed ratios and Mach numbers. The details of the apparatus have been covered in several prior studies \citep{miller_horizontal_2019,pique_dominant_2022,pique_laboratory_2022,kurelek_performance_2023,pique_understanding_2025}, so only a brief summary is provided here. A schematic of the experiment is shown in Fig.\ \ref{fig:setup}.

All tests in this study were performed at $Re_D = 4\times 10^6$, using a range of static pressures of up to 170 bar and free-stream velocities between 3.0 and 3.6 $\rm{ms^{-1}}$ to keep $Re_D$ constant to within $\pm 3\%$. The ambient temperature and pressure in the test section were continuously logged during experiments, and the free-stream velocity was measured using a pitot-static tube located $5.1D$ upstream of the test stand.

A self-starting, three-bladed wind turbine, identical to that employed by \citet{miller_horizontal_2019} except scaled to a rotor diameter of $D = 15$ cm, was used in the experiments. The total blockage when accounting for the rotor swept area and tower was 11\%. Aerodynamic forces were measured using a six-axis load cell (JR3 Inc.\ 30E12A4, 100 N range) at a sampling rate of 1 kHz. Torque and rotation rate were measured using a torque transducer (Magtrol TM-303, 0.5 Nm range) sampled at 200 kHz, from which power was calculated. The load on the turbine was set using a magnetic hysteresis brake (Magtrol AHB-3). Turbine thrust, power, and tip-speed ratio measurements were corrected for blockage using the method of \citet{bahaj_power_2007}, and power measurements were also corrected for gearbox losses following \citet{miller_horizontal_2019}. Average measurement uncertainties, quantified using the mean standard error over one turbine rotation, are 2.7\% for $C_t$ and 0.88\% for $C_p$.

To achieve time-varying perturbations in the wake of the turbine, the braking load applied to the turbine rotor was varied sinusoidally via  a reference signal provided to the brake through an amplifier (Kepco BOP72-1.5M). This input waveform was recorded alongside the turbine force and power data for phase synchronization across measurement ensembles.

Measurements of the streamwise flow velocity were obtained in the wake of the turbine by means of a nano-scale thermal anemometry probe (NSTAP) \citep{bailey_turbulence_2010,fan_nanoscale_2015} and constant-temperature anemometry bridge (Dantec StreamLine Pro) at a sampling rate of 200 kHz. The probe was positioned using a three-axis traverse mounted inside the test section. Two types of measurement sweeps were conducted: (i) streamwise sweeps of 100 logarithmically spaced points along the turbine centerline ($r=0$) over $1.44 \leq x/D \leq 10.7$, and (ii) radial sweeps at $x/D=1.50$, 3.50, and 7.00 over $-1.18 \leq r/D \leq 1.18$. Calibrations were performed with a reference pitot-static tube before and after each measurement sweep to account for temperature drift in the tunnel. The mean standard error of the phase-averaged velocities is between 1.4\% and 1.8\%, depending on the measurement locations relative to the wake.

\subsection{Design of experiments}\label{sec:methods_design}

The aim of the experiments is to evaluate the effects of wake advection on time-varying wake dynamics. To this end, slow time variations in the turbine rotation rate were chosen as a method for forcing the wake in a nominally quasi-steady manner. Forcing the wake at higher frequencies has been shown to excite nonlinear wake dynamics that are generally only present when active control is applied \citep{meyers_wind_2022,messmer_enhanced_2024,wei_wake_2024}, and not in the slower time-varying disturbances relevant to typical wind-farm operation. Practically, slow time variations also allowed for large-amplitude oscillations in tip-speed ratio, which improved the signal-to-noise ratio of the data while still avoiding nonlinearities. Thus, for these experiments, the Strouhal number of the tip-speed ratio variations, defined as 

\begin{equation}
    St = \frac{fD}{U_\infty},
\end{equation}

\noindent was kept in the range $0.02\leq St \leq 0.06$ ($0.50 \leq f \leq 1.52$ Hz with $3.0 \leq U_\infty \leq 3.6$ $\rm{ms^{-1}}$). This is well below typical values for dynamic induction control, $St\gtrsim 0.2$ \citep[\textit{cf.},][]{messmer_enhanced_2024}, and therefore is expected to be well within the linear dynamical regime.

\begin{figure}
\begin{subfigure}[t]{0.48\textwidth}
\centering
  \includegraphics[width=\textwidth]{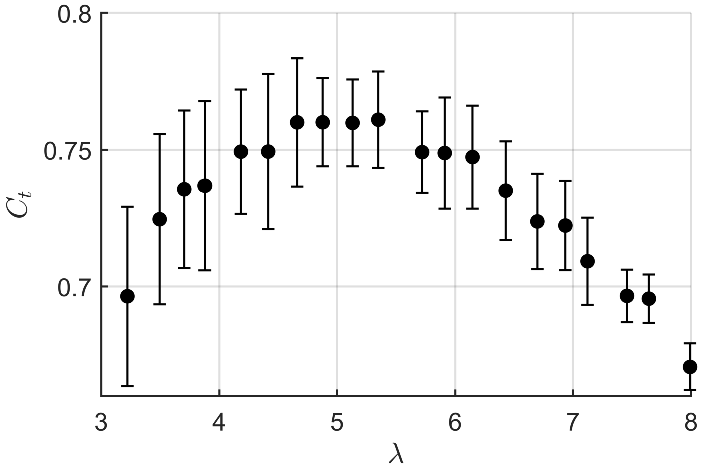}
  \caption{}
\label{fig:steady_Ct}
\end{subfigure}
\hfill
\begin{subfigure}[t]{0.48\textwidth}
\centering
  \includegraphics[width=\textwidth]{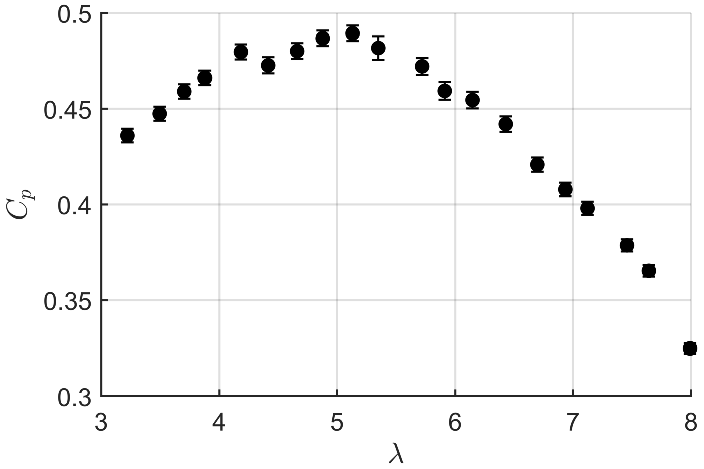}
  \caption{}
\label{fig:steady_Cp}
\end{subfigure}
\caption{Steady-flow coefficient of thrust (a) and coefficient of power (b).}
\label{fig:steady_curves}
\end{figure}

Test cases were selected based on the steady-flow thrust and power curves of the turbine, shown in Fig.\ \ref{fig:steady_curves}. The thrust curve has a parabolic shape, with a peak at $\lambda_\text{opt} \approx 5$. Therefore, assuming quasi-steady variations along the thrust curve, the local relationship of $C_t$ and $\lambda$ changes depending on the operating point. For $\lambda<\lambda_\text{opt}$, $C_t$ increases with $\lambda$. For $\lambda>\lambda_\text{opt}$, $C_t$ decreases with increasing $\lambda$. Thus, by operating at different mean tip-speed ratios, $\overline{\lambda}$, and imposing variations with peak amplitude $\hat{\lambda}$, different time-varying forcing combinations of thrust and tip-speed ratio could be applied to the wake.

The main experiments presented here involved variations in tip-speed ratio with an amplitude of $\hat{\lambda} = 0.9$ at a Strouhal number of $St=0.04$. Three mean loading, and therefore steady-flow, conditions were investigated: $\overline{\lambda}=3.87 \pm 0.07$, $4.89 \pm 0.09$, and $5.86 \pm 0.04$.  For the time-varying conditions, the brake and turbine gearbox exhibited some hysteresis as a function of changing temperature and pressure conditions in the HRTF. As such, iterative adjustments were made to the mean and peak braking loads prior to each measurement series to ensure $\overline{\lambda}$ and $\hat{\lambda}$ variations remained within predetermined limits of $\overline{\lambda} \pm 0.1$ and $\hat{\lambda} \pm 0.03$. For simplicity, throughout the remaining discussion both the steady and time-varying cases are denoted by $\lambda = \overline{\lambda} + \hat{\lambda}$, with $\hat{\lambda} \approx 0$ assumed for the steady cases, and all other values rounded to one significant digit, thus denoting the time-varying cases as $\overline{\lambda} = 4$, 5 and 6 with $\pm \hat{\lambda} = 0.9$. More precise values of $\lambda$ for each test are given in figure legends and titles for the sake of completeness.

For each case, wake measurements were recorded in streamwise and radial sweeps as described in \S\ref{sec:methods_apparatus}. At least 30 full forcing cycles were recorded at each measurement location. The tunnel was allowed to settle for at least 20 minutes between sweeps to allow the temperature to stabilize and thereby ensure accurate calibrations.


Data from time-varying cases were phase-averaged and synchronized based on the phase of the tip-speed ratio waveform. Phase-averaged velocities are denoted by $U$ and time-averaged quantities with an overline, \textit{e.g.}, $\overline{U}$, with the exception of the free-stream velocity $U_\infty$, which is assumed to be constant. The quantity ${u'}^2$ represents the streamwise-velocity variance at a given point and phase, taken over the ensemble of cycles used for phase-averaging. The quantity $\overline{{u'}^2}$ is the time-averaged streamwise-velocity variance.

Finally, two additional forcing frequencies ($St=0.02$ and 0.06) and tip-speed ratio amplitudes ($\hat{\lambda}=0.4$ and 1.5) were investigated for $\overline{\lambda}\approx 5$. These results are presented in App.\ \ref{app:St} and \ref{app:amp}, respectively.

%% file: s4_results.tex
\section{Results}\label{sec:results}

\subsection{Steady-flow wake measurements}\label{sec:results_steady}

\begin{figure}
\begin{subfigure}[t]{0.48\textwidth}
\centering
  \includegraphics[width=\textwidth]{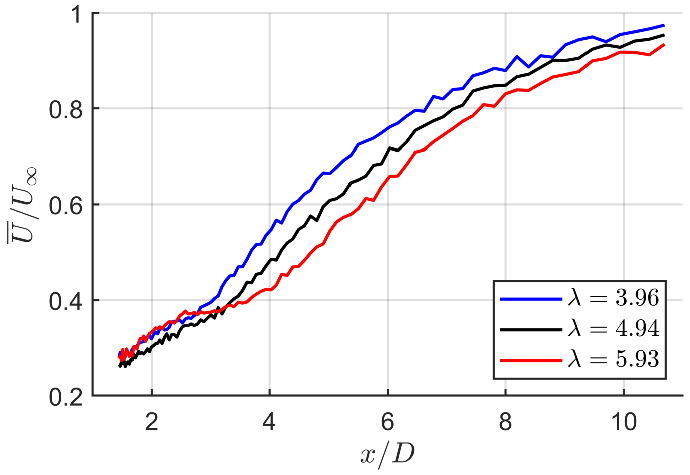}
  \caption{}
\label{fig:steady_U}
\end{subfigure}
\hfill
\begin{subfigure}[t]{0.48\textwidth}
\centering
  \includegraphics[width=\textwidth]{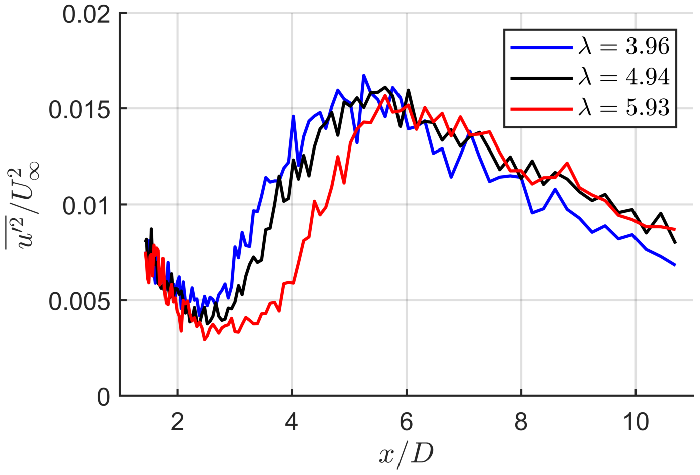}
  \caption{}
\label{fig:steady_Uvar}
\end{subfigure}
\caption{Time-averaged streamwise velocity (a) and velocity variance (b) for three tip-speed ratios in steady flow.}
\label{fig:steady_x}
\end{figure}

Steady-flow measurements for $\overline{\lambda}\approx4$, 5, and 6 are presented here as a reference for the time-varying cases. The observed trends are similar to those reported by \citet{pique_understanding_2025}. Velocity and velocity-variance measurements along the wake centerline for the three cases are shown in Fig.\ \ref{fig:steady_x}. Velocities in the near-wake region ($x/D \lesssim 3$) are similar for all three cases. However, the streamwise extent of the near-wake region increases monotonically with tip-speed ratio, indicated by the downstream shifts in the location at which the mean velocity profile steepens and where the velocity variance begins to increase, both occurring within  3 $\lesssim x/D \lesssim 4$ in figures \ref{fig:steady_U} and \ref{fig:steady_Uvar}, respectively. This is likely due to the helical tip-vortex system persisting further downstream, pushing the point at which the intermediate-wake shear layers converge further downstream.

\begin{figure}
\begin{subfigure}[t]{0.32\textwidth}
\centering
  \includegraphics[width=\textwidth]{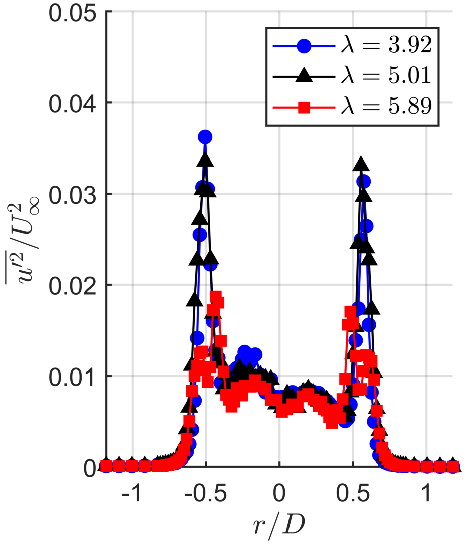}
  \caption{}
\label{fig:steady_Uvar_x1.50D}
\end{subfigure}
\hfill
\begin{subfigure}[t]{0.32\textwidth}
\centering
  \includegraphics[width=\textwidth]{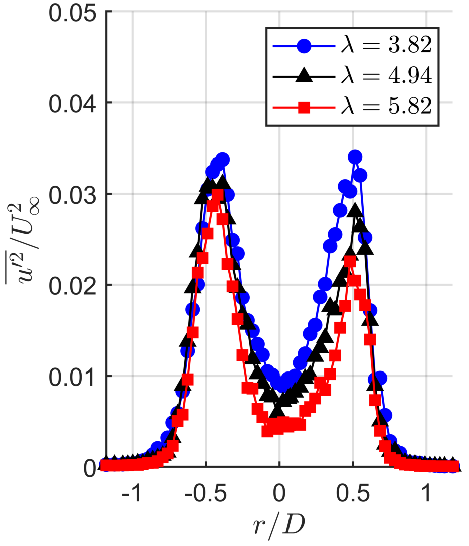}
  \caption{}
\label{fig:steady_Uvar_x3.50D}
\end{subfigure}
\hfill
\begin{subfigure}[t]{0.32\textwidth}
\centering
  \includegraphics[width=\textwidth]{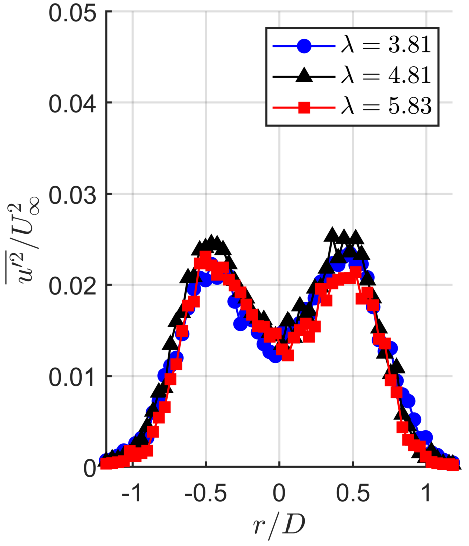}
  \caption{}
\label{fig:steady_Uvar_x7.00D}
\end{subfigure}
\caption{Radial variations in time-averaged streamwise-velocity variance at $x/D=1.50$ (a), 3.50 (b), and 7.00 (c) at three tip-speed ratios in steady flow.}
\label{fig:steady_r}
\end{figure}

Radial sweeps of the streamwise velocity variance are shown in Fig.\ \ref{fig:steady_r} for three downstream locations. At $x/D=1.50$ (Fig.\ \ref{fig:steady_Uvar_x1.50D}), the peaks in the velocity variance around $r/D=\pm0.5$ are relatively high, indicating relatively high strength and coherence of the tip vortices at this station. At $x/D=3.50$ (Fig.\ \ref{fig:steady_Uvar_x3.50D}), the peaks have broadened and reduced in amplitude, indicating the breakdown of the tip vortices and growth in the shear layers from the near wake to the far wake. In the far wake ($x/D=7.00$, Fig.\ \ref{fig:steady_Uvar_x7.00D}), no significant differences are evident as a function of tip-speed ratio. The differences in wake recovery seen in figures \ref{fig:steady_x} and \ref{fig:steady_r} suggest that the breakdown of the near wake sets the initial conditions for the self-similar development of the far wake. This implies that varying the tip-speed ratio in time will dynamically shift the near- and intermediate-wake boundaries, thus changing the overall wake structure.

\subsection{Time-varying wake dynamics}\label{sec:results_time-varying}

\begin{figure}
\begin{subfigure}[t]{0.48\textwidth}
\centering
  \includegraphics[width=\textwidth]{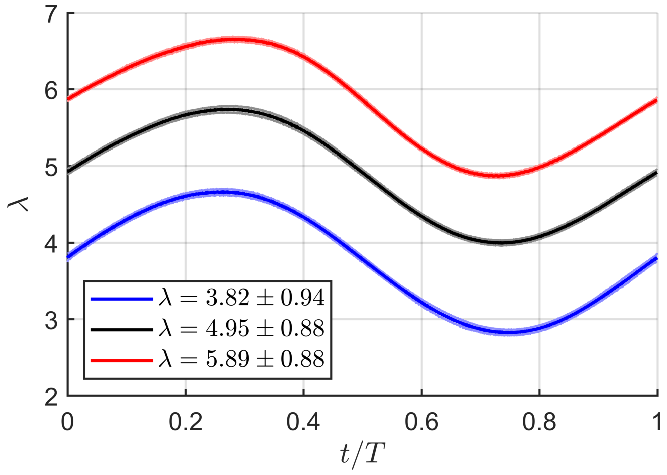}
  \caption{}
\label{fig:TSR_varyTSR}
\end{subfigure}
\hfill
\begin{subfigure}[t]{0.48\textwidth}
\centering
  \includegraphics[width=\textwidth]{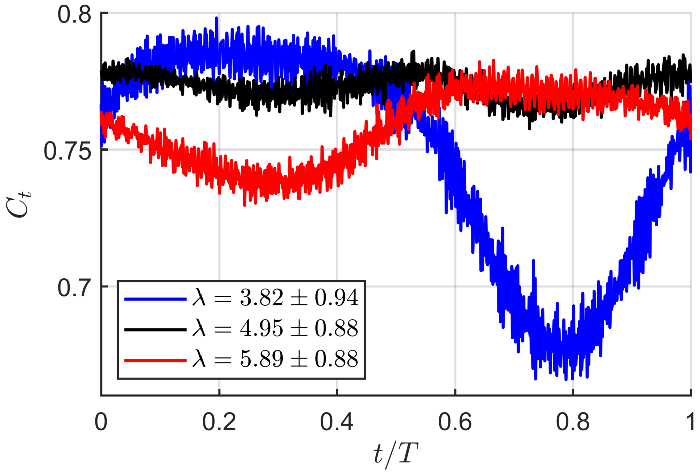}
  \caption{}
\label{fig:Ct_varyTSR}
\end{subfigure}
\caption{Phase-averaged tip-speed ratio (a) and thrust coefficient (b) for three mean tip-speed ratios. The $\pm$ symbol in the plot legends denotes the tip-speed ratio amplitude $\hat{\lambda}$, not an uncertainty. For all cases, $St = 0.04$.}
\label{fig:Ct_TSR}
\end{figure}

For the time-varying experiments, phase-averaged profiles of tip-speed ratio and thrust coefficient with respect to the intracycle phase, $t/T$ (where $T=1/f$), are shown in Fig.\ \ref{fig:Ct_TSR} for the three tested cases. Fig.\ \ref{fig:Ct_varyTSR} establishes that the variations in $C_t$ are nominally quasi-steady, as across all cases, $C_t$ at any particular phase can be recovered from the steady-flow thrust curve (Fig.\ \ref{fig:steady_Ct}) in accordance with the $\lambda$ value at that phase. For the $\overline{\lambda}\approx4$ case, $C_t$ increases with $\lambda$, and so the phase-averaged thrust coefficient is in phase with the tip-speed ratio oscillation. Conversely, for $\overline{\lambda}\approx6$, the time-varying $C_t$ profile is out of phase with $\lambda$, due to the negative slope of the thrust curve for $\lambda > \lambda_\text{opt}$. The $\overline{\lambda}\approx5$ case shows slight variations, with $C_t$ remaining constant within the measurement uncertainty as the thrust curve is relatively flat about $\lambda=5$. 

\begin{figure}
\begin{subfigure}[t]{0.48\textwidth}
\centering
  \includegraphics[width=\textwidth]{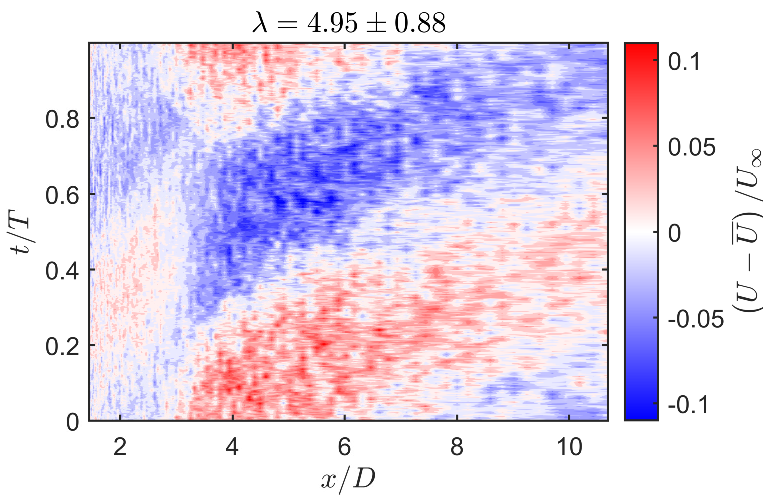}
  \caption{}
\label{fig:xt_Uprime_TSR5}
\end{subfigure}
\hfill
\begin{subfigure}[t]{0.48\textwidth}
\centering
  \includegraphics[width=\textwidth]{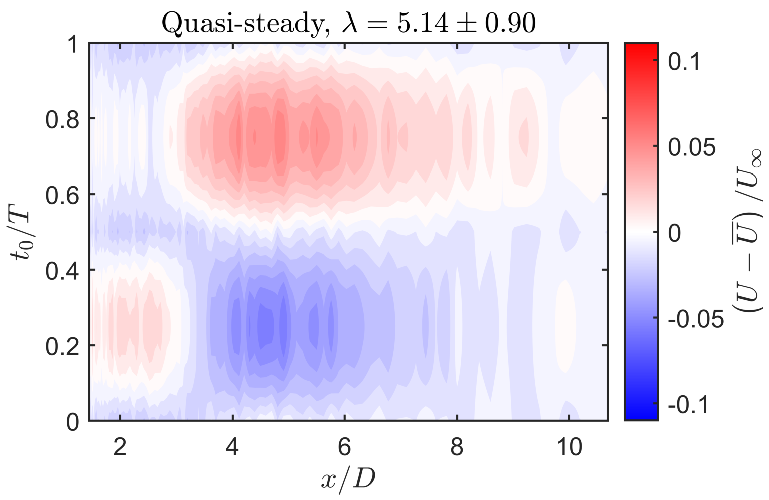}
  \caption{}
\label{fig:xt_Uprime_TSR5_QS}
\end{subfigure}
\centering\begin{subfigure}[t]{0.48\textwidth}
\centering
  \includegraphics[width=\textwidth]{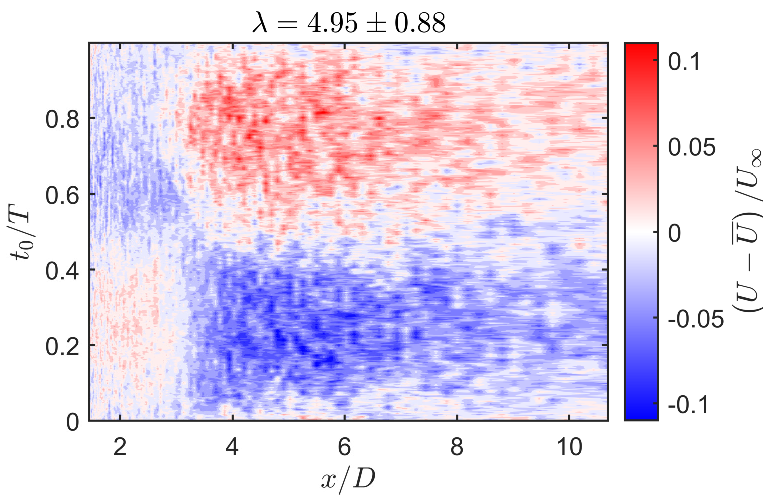}
  \caption{}
\label{fig:xt0_Uprime_TSR5}
\end{subfigure}
\caption{Variations in the streamwise-velocity perturbation, $\tilde{u}/U_\infty = (U-\overline{U})/U_\infty$, along the wake centerline for $\overline{\lambda} \approx 5$ and $\hat{\lambda}\approx 0.9$. Phase-averaged data for $St = 0.04$ are shown in (a) over a single period $T$, while a quasi-steady solution interpolated from the steady-flow data in Fig.\ \ref{fig:steady_U} is shown in (b). The same phase-averaged data from (a) are shown after a Lagrangian transformation to remove the traveling-wave velocity in (c), plotted along characteristics defined by an initial phase $t_0/T$.}
\label{fig:xt_Uprime_PAvQS}
\end{figure}

In Fig.\ \ref{fig:xt_Uprime_TSR5}, the phase-averaged centerline wake velocities are plotted as contours in a space spanned by streamwise distance, $x/D$, and intracycle phase, $t/T$. The contour levels present the normalized phase-averaged amplitude of the velocity perturbations, $\tilde{u}/U_\infty = (U(x,t) - \overline{U}(x))/U_\infty$. These data thus highlight the evolution of velocity perturbations in time and along the wake centerline in the streamwise direction. They are compared against quasi-steady reconstructions of the wake evolution, shown in Fig.\ \ref{fig:xt_Uprime_TSR5_QS}. These reconstructions are derived from linear interpolation of the steady-flow wake measurements (figures \ref{fig:steady_x} and \ref{fig:steady_r}) according to the measured time-varying profile of the tip-speed ratio about a mean value of $\overline{\lambda} \approx 5$ (black curve in Fig.\ \ref{fig:TSR_varyTSR}). Physically, these quasi-steady reference solutions are most applicable when the wake forcing time scales are much slower than the propagation time scales of disturbances through the wake, as true quasi-steady behavior would require the wake to respond globally and instantaneously to changes at the rotor plane.

Traveling waves are evident in the data from the time-varying experiment shown in Fig.\ \ref{fig:xt_Uprime_TSR5}, seen as red and blue streaks that travel along curved lines in the $x-t$ plane. These curved trajectories show that the velocity perturbations do not propagate along the wake centerline at a constant velocity, but rather travel at a slower velocity in the near wake and accelerate into the far wake. Since nonlinearly advecting traveling waves are a time-varying phenomenon, the wake does not behave in a quasi-steady manner, with the reference solution given in Fig.\ \ref{fig:xt_Uprime_TSR5_QS} exhibiting entirely different propagation characteristics that are not indicative of traveling waves.

The advective nature of the wake demonstrated here affirms the theoretical analysis given in \S\ref{sec:theory}, showing empirically that information must propagate along characteristic curves in $x-t$ space. It is thus beneficial to plot the data in terms of a phase of origin, $t_0/T$, to highlight the time evolution of fluid parcels as they advect downstream. This is done for each phase-averaged streamwise dataset by forward-Euler integration over the velocity fields. For a fluid parcel at a point $(x_k,t_k)$, the point it would occupy in the next time step is given by $(x_{k+1}, t_{k+1})$ where $x_{k+1} = x_k + U(x_k,t_k)\Delta t$ and $t_{k+1} = t_k + \Delta t$. By starting at $t_0/T \in [0,1)$ and $x/D=1.44$, and interpolating from the phase-averaged velocity fields $U(x,t)$ as needed, Lagrangian wake-following contours are constructed in terms of the characteristic variable $t_0$ and streamwise distance. This process effectively flattens the characteristic curves from Fig.\ \ref{fig:xt_Uprime_TSR5} onto horizontal lines with constant $t_0/T$, as shown in Fig.\ \ref{fig:xt0_Uprime_TSR5}. These Lagrangian-transformed contours recover the profiles of the quasi-steady reference solution (Fig.\ \ref{fig:xt_Uprime_TSR5_QS}). The uniform collapse of the data onto horizontal lines under this Lagrangian transformation emphasizes the importance of traveling waves for these time-varying wake dynamics. Furthermore, it demonstrates that advection occurs at the wake velocity and not the free-stream velocity, as a spatially varying advection velocity is required to account for the curved trajectories in the original phase-averaged data.

\begin{figure}
\begin{subfigure}[t]{0.48\textwidth}
\centering
  \includegraphics[width=\textwidth]{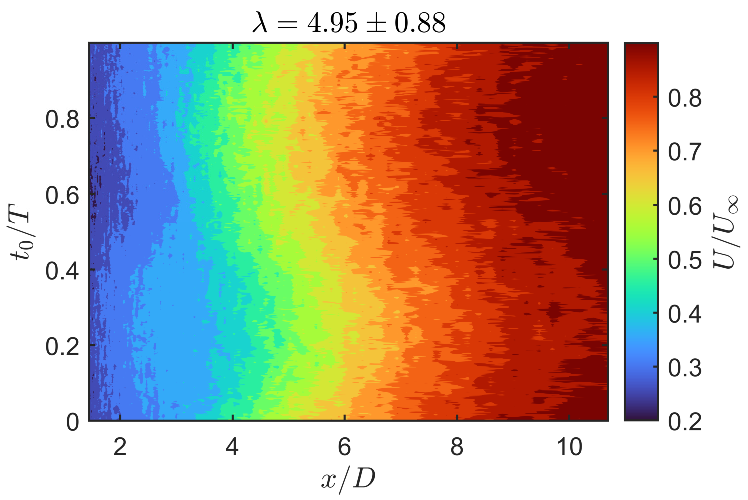}
  \caption{}
\label{fig:xt0_U_TSR5}
\end{subfigure}
\hfill
\begin{subfigure}[t]{0.48\textwidth}
\centering
  \includegraphics[width=\textwidth]{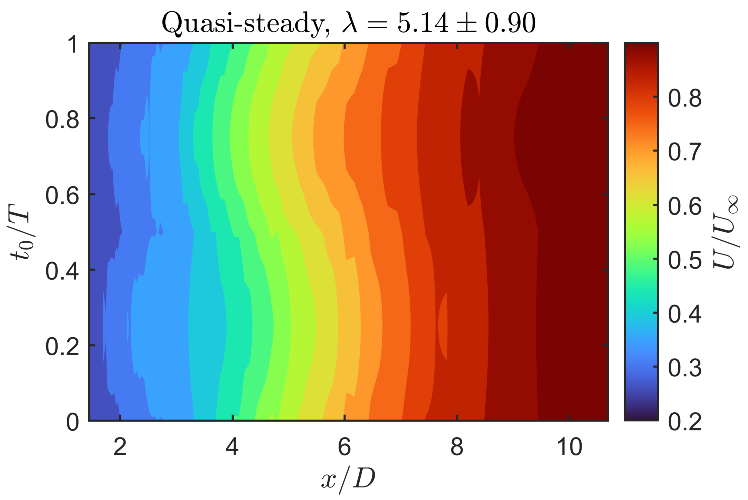}
  \caption{}
\label{fig:xt0_U_TSR5_QS}
\end{subfigure}
\caption{Phase-averaged variations in the streamwise velocity along the wake centerline for $\overline{\lambda}\approx5$ (a), and the interpolated quasi-steady reference solution (b), plotted along characteristics defined by an initial phase $t_0/T$. For both cases, $\hat{\lambda}\approx 0.9$.}
\label{fig:xt0_U_PAvQS}
\end{figure}

\begin{figure}
\begin{subfigure}[t]{0.48\textwidth}
\centering
  \includegraphics[width=\textwidth]{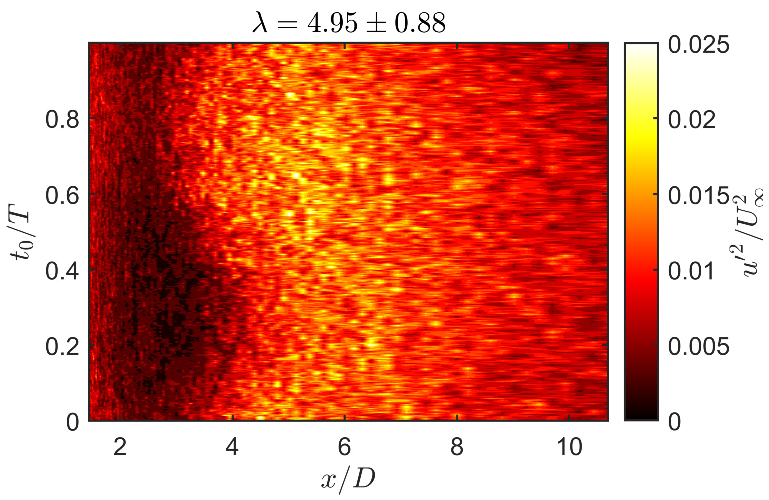}
  \caption{}
\label{fig:xt0_Uvar_TSR5}
\end{subfigure}
\hfill
\begin{subfigure}[t]{0.48\textwidth}
\centering
  \includegraphics[width=\textwidth]{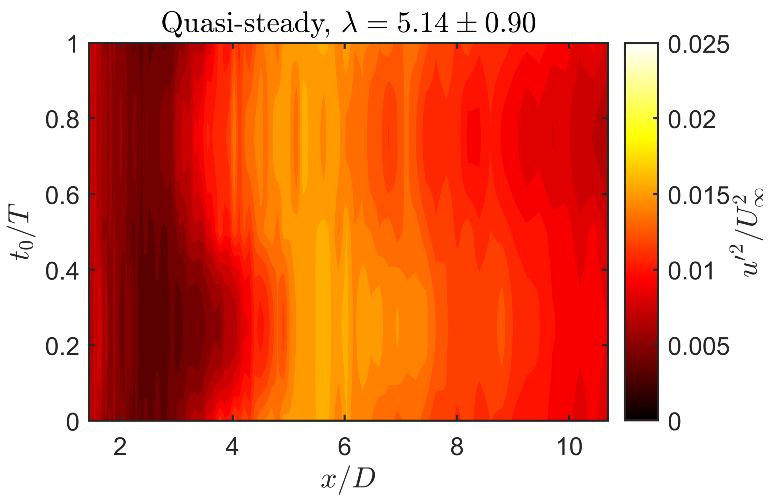}
  \caption{}
\label{fig:xt0_Uvar_TSR5_QS}
\end{subfigure}
\caption{Phase-averaged variations in the streamwise-velocity variance along the wake centerline for $\overline{\lambda}\approx5$ (a), and the interpolated quasi-steady reference solution (b), plotted along characteristics defined by an initial phase $t_0/T$. For both cases, $\hat{\lambda}\approx 0.9$.}
\label{fig:xt0_Uvar_PAvQS}
\end{figure}

To underscore these points, profiles of the phase-averaged velocity, $U$, and velocity variance, $u'^2$, for $\overline{\lambda}\approx 5$ and an amplitude of $\hat{\lambda}\approx 0.9$ are plotted as a function of the characteristic time of origin $t_0$ and compared with the corresponding quasi-steady reference solutions in figures \ref{fig:xt0_U_PAvQS} and \ref{fig:xt0_Uvar_PAvQS}, respectively. The Lagrangian-transformed time-varying data closely correspond to the quasi-steady solutions, further demonstrating that quasi-steady wake dynamics are recovered at these low forcing frequencies only after wake advection has been taken into account.

It is important to note that, since data were not recorded for $x/D<1.44$, the inception of disturbances from the rotor plane cannot be identified. However, as noted in \S\ref{sec:theory_waves}, pressure recovery often occurs in the region of the near wake immediately downstream of the rotor plane. In this region, pressure-gradient forces are significant, and since these forces are nonlocal \citep{bempedelis_analytical_2022}, thrust information from the turbine may be communicated nearly instantaneously downstream until the pressure gradient decreases in magnitude. If true, this would imply that traveling waves produced by thrust disturbances will appear to begin propagating not from $x/D=0$ but from farther downstream, where the decreasing influence of pressure-gradient forces allows disturbances to travel at finite speeds. Additional measurements in the near wake would be required to verify this conjecture.

\subsection{Effects of turbine operation on time-varying wake dynamics}\label{sec:results_TSR-varying}

\begin{figure}
\begin{subfigure}[t]{0.48\textwidth}
\centering
  \includegraphics[width=\textwidth]{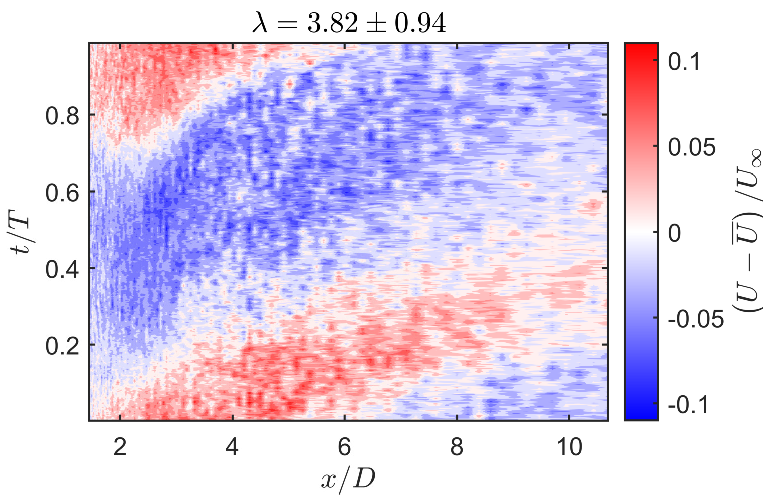}
  \caption{}
\label{fig:xt_Uprime_TSR4}
\end{subfigure}
\hfill
\begin{subfigure}[t]{0.48\textwidth}
\centering
  \includegraphics[width=\textwidth]{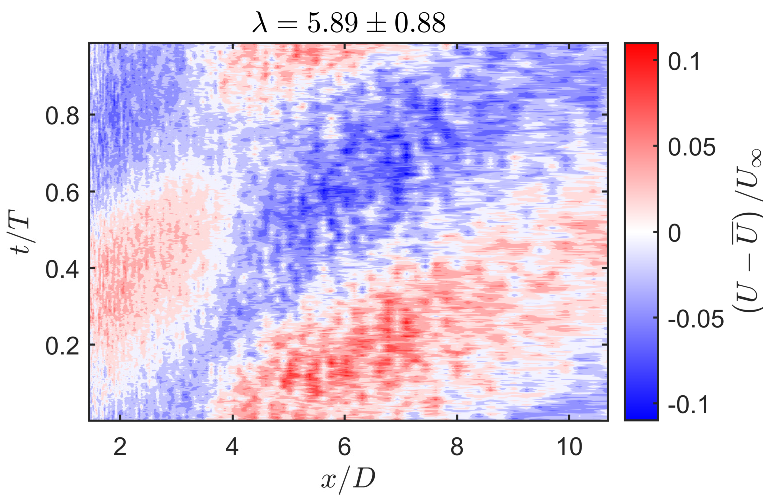}
  \caption{}
\label{fig:xt_Uprime_TSR6}
\end{subfigure}
\caption{Phase-averaged variations in the streamwise-velocity perturbation, $\tilde{u}/U_\infty = (U-\overline{U})/U_\infty$, along the wake centerline for $\overline{\lambda} \approx 4$ (a) and $6$ (b). For both cases, $\hat{\lambda}\approx 0.9$ and $St = 0.04$.}
\label{fig:xt_Uprime_4v6}
\end{figure}

\begin{figure}
\begin{subfigure}[t]{0.48\textwidth}
\centering
  \includegraphics[width=\textwidth]{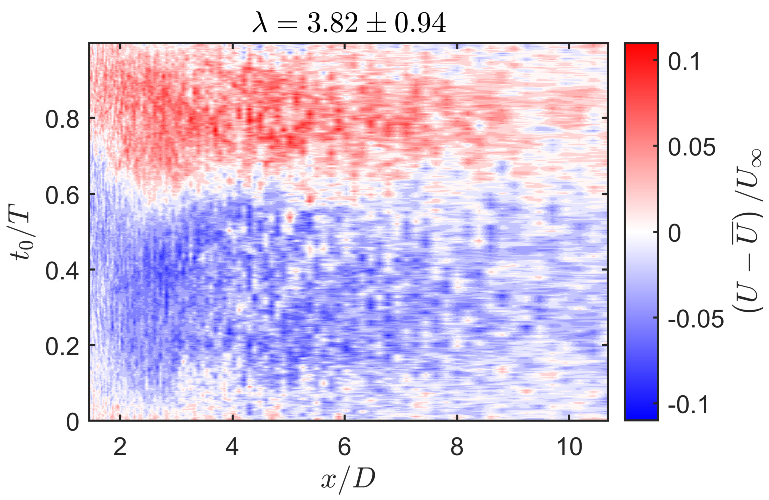}
  \caption{}
\label{fig:xt0_Uprime_TSR4}
\end{subfigure}
\hfill
\begin{subfigure}[t]{0.48\textwidth}
\centering
  \includegraphics[width=\textwidth]{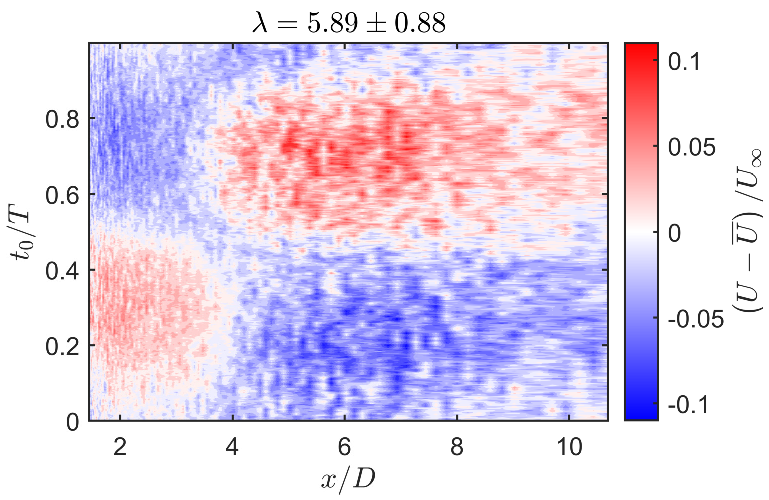}
  \caption{}
\label{fig:xt0_Uprime_TSR6}
\end{subfigure}
\caption{Phase-averaged variations in the streamwise-velocity perturbation along the wake centerline for $\overline{\lambda}\approx4$ (a) and $6$ (b), plotted along characteristics defined by an initial phase $t_0/T$. For both cases, $\hat{\lambda}\approx 0.9$ and $St = 0.04$.}
\label{fig:xt0_Uprime_4v6}
\end{figure}

Given that the time-varying wake structure can be described in a quasi-steady manner after accounting for wake advection, we now explore how quasi-steady turbine operation can be employed to manipulate the wake structure. This is examined through cases of time-varying tip-speed ratio about mean values of $\overline{\lambda}\approx4$ and 6. As shown in Fig.\ \ref{fig:Ct_TSR}, imposing periodic rotation-rate oscillations at $\overline{\lambda}\approx4$ produces thrust variations that are in phase with $\lambda(t)$, while at $\overline{\lambda}\approx6$ the thrust variations are out of phase with $\lambda(t)$. The phase-averaged velocity perturbations for these two cases are shown in Fig.\ \ref{fig:xt_Uprime_4v6}. As for the case of $\overline{\lambda}\approx5$, curved trajectories indicative of nonlinearly advecting velocity-perturbation waves are evident in these cases. While no steady-flow measurements are available at $\overline{\lambda}\approx3$ and 8 to construct quasi-steady reference solutions, Lagrangian wake transformations do collapse the traveling waves onto horizontal lines in $x-t_0$ space (Fig.\ \ref{fig:xt0_Uprime_4v6}). This suggests that these wake dynamics can also be considered quasi-steady after accounting for wake advection.


Though the tip-speed ratio amplitudes are identical and the mean thrust coefficients are similar for these two cases ($\overline{C_t}\approx0.75$, \textit{cf}., Fig.\ \ref{fig:steady_Ct}) and in spite of the quasi-steady nature of the wake dynamics, differences in wake structure are still apparent. For $\overline{\lambda}\approx4$, in which thrust and tip-speed ratio vary in phase with each other, the waves advect downstream through the wake unimpeded, indicated by the continuous red and blue swaths in Fig.\ \ref{fig:xt0_Uprime_TSR4}. By contrast, in the cases of both $\overline{\lambda}\approx5$ and 6, the waves abruptly change sign at $x/D\approx 3$ to 4 (figures \ref{fig:xt0_Uprime_TSR5} and \ref{fig:xt0_Uprime_TSR6}, respectively).

\begin{figure}
\begin{subfigure}[t]{0.48\textwidth}
\centering
  \includegraphics[width=\textwidth]{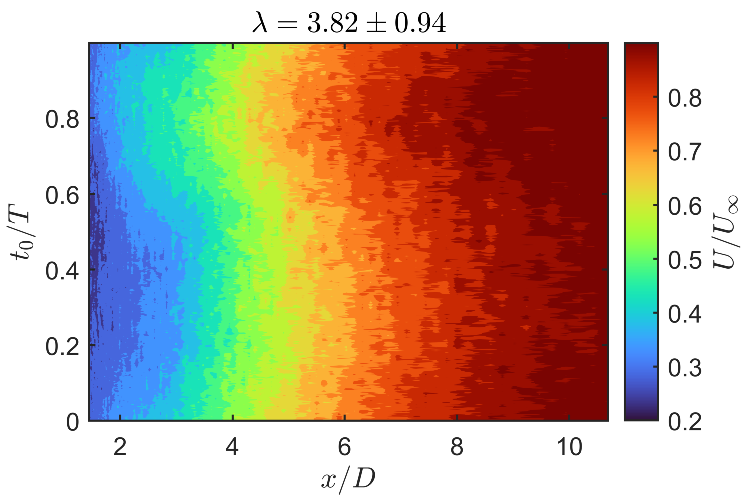}
  \caption{}
\label{fig:xt0_U_TSR4}
\end{subfigure}
\hfill
\begin{subfigure}[t]{0.48\textwidth}
\centering
  \includegraphics[width=\textwidth]{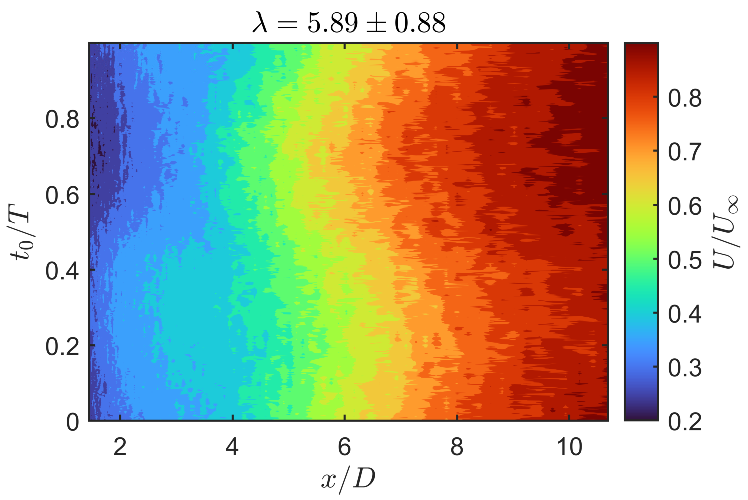}
  \caption{}
\label{fig:xt0_U_TSR6}
\end{subfigure}
\caption{Phase-averaged streamwise velocities along the wake centerline for $\overline{\lambda}\approx4$ (a) and $6$ (b), plotted along characteristics defined by an initial phase $t_0/T$. For both cases, $\hat{\lambda}\approx 0.9$ and $St = 0.04$.}
\label{fig:xt0_U_4v6}
\end{figure}

To explain this wave inversion, Lagrangian-transformed maps of the phase-averaged streamwise velocity $U$ are plotted for $\overline{\lambda}\approx 4$ and $6$ in Fig.\ \ref{fig:xt0_U_4v6}. For $\overline{\lambda}\approx4$ (Fig.\ \ref{fig:xt0_U_TSR4}), the entire wake shifts back and forth in the streamwise direction -- highlighted by considering the highest velocity-deficit region indicated by the blue contours, which recede upstream as the phase advances, creating S-shaped contours. In contrast, the other two cases (figures \ref{fig:xt0_U_TSR5} and \ref{fig:xt0_U_TSR6}) show the intermediate-wake region ($2\lesssim x/D \lesssim 6$) is deformed by an alternate expansion (for $t_0/T < 0.5$) and compression (for $t_0/T > 0.5$) in the streamwise direction. Such wake deformations can be linked to the phase inversions seen in figures \ref{fig:xt0_Uprime_TSR5} and \ref{fig:xt0_Uprime_TSR6}. By definition, the time-averaged velocity profile $\overline{U}$ is not a function of time. Additionally, we note that the instantaneous velocity also remains relatively constant in time at the center of the deformed region ($x/D\approx4$). Thus, when the wake expands ($t_0/T < 0.5$), the slope of the velocity profile decreases, which acting about a central location of $x/D \approx 4$, yields instantaneous velocities that are higher upstream and lower downstream than the corresponding time-averaged values. This will result in a change in sign of the velocity perturbation across this location. When the wake is compressed in the following half-cycle, the slope of the velocity profile increases, resulting in the opposite effect on the velocity-perturbation profile. The phase inversion of the traveling waves across $x/D\approx 4$ is therefore a consequence of these deformations in the wake.

\begin{figure}
\begin{subfigure}[t]{0.48\textwidth}
\centering
  \includegraphics[width=\textwidth]{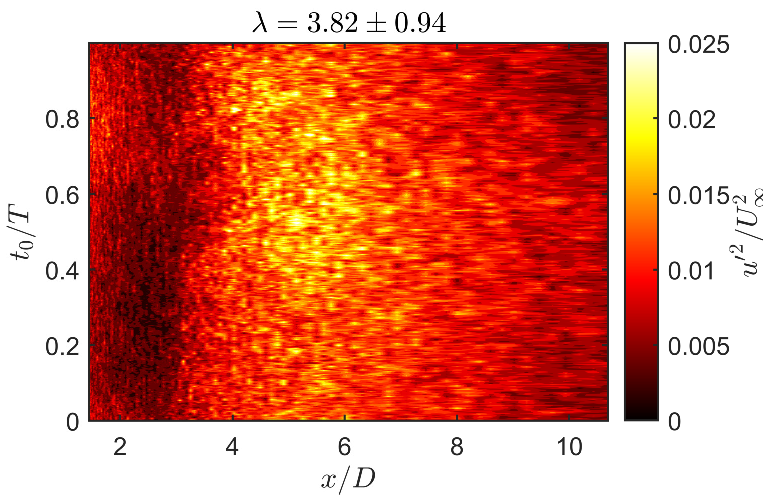}
  \caption{}
\label{fig:xt0_Uvar_TSR4}
\end{subfigure}
\hfill
\begin{subfigure}[t]{0.48\textwidth}
\centering
  \includegraphics[width=\textwidth]{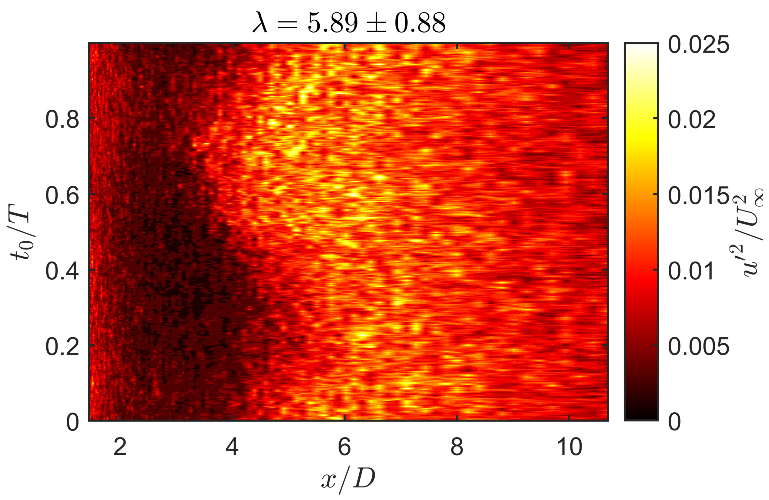}
  \caption{}
\label{fig:xt0_Uvar_TSR6}
\end{subfigure}
\caption{Phase-averaged variations in the streamwise-velocity variance along the wake centerline for $\overline{\lambda}\approx4$ (a) and $6$ (b), plotted along characteristics defined by an initial phase $t_0/T$. For both cases, $\hat{\lambda}\approx 0.9$ and $St = 0.04$.}
\label{fig:xt0_Uvar_4v6}
\end{figure}

We argue that the deformations in the intermediate wake result from changes in the tip-vortex breakdown location. Such effects are more readily evident in Lagrangian-transformed contours of the streamwise-velocity variance ${u'}^2$, presented in Fig.\ \ref{fig:xt0_Uvar_4v6}. In these plots, the near-wake region appears as a dark region with low variance, since data are recorded along the wake centerline. When the helical tip-vortex system breaks down, the shear layers bounding the wake converge on the centerline, and the velocity variance peaks locally. The tip-vortex breakdown location can thus be visualized by the sharp transition between the dark and light regions. For $\overline{\lambda}\approx4$ (Fig.\ \ref{fig:xt0_Uvar_TSR4}), the tip-vortex breakdown location is farthest upstream when the tip-speed ratio is highest and farthest downstream when $\lambda$ is lowest, which occurs for $t/T < 0.5$ and $t/T > 0.5$ respectively (\textit{cf}., Fig.\ \ref{fig:TSR_varyTSR}). These shifts are relatively small since, as evident in the steady-flow variance profiles (Fig.\ \ref{fig:steady_Uvar}), the tip-vortex breakdown location does not change appreciably between steady tip-speed ratios of $\overline{\lambda} = 4$ and 5. In contrast, the higher tip-speed-ratio cases ($\overline{\lambda}\approx5$ and 6) show the opposite behavior, with the tip-vortex breakdown location moving downstream when $\lambda$ is the highest and vice versa (figures \ref{fig:xt0_Uvar_TSR5} and \ref{fig:xt0_Uvar_TSR6}). Thus, for these cases, the tip-vortex breakdown location moves in phase with $\lambda$.

\begin{figure}
\centering
\begin{subfigure}[t]{0.48\textwidth}
  \includegraphics[width=\textwidth]{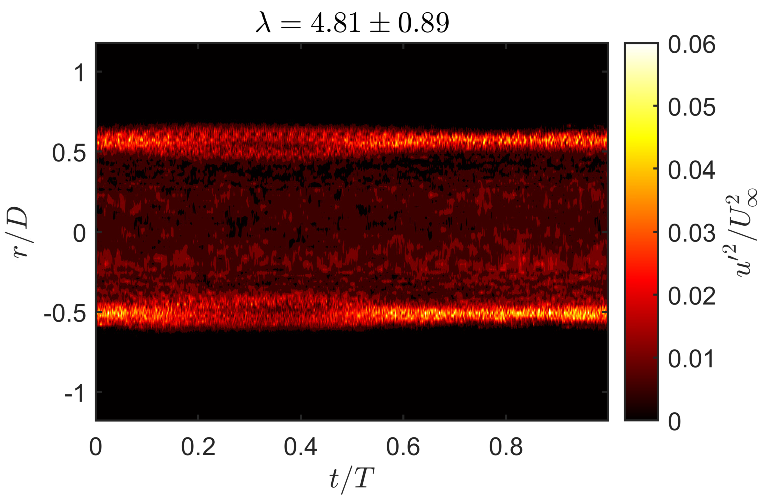}
  \caption{}
\label{fig:rt1-5_Uvar}
\end{subfigure}
\begin{subfigure}[t]{0.48\textwidth}
  \includegraphics[width=\textwidth]{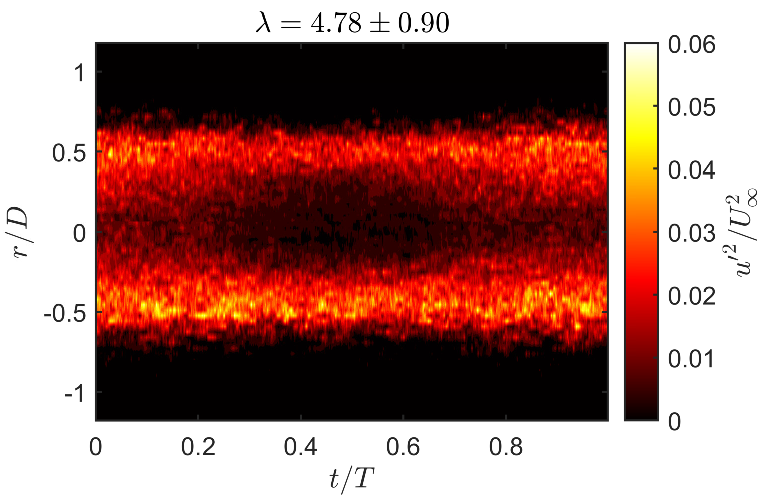}
  \caption{}
\label{fig:rt3-5_Uvar}
\end{subfigure}
\caption{Phase-averaged streamwise-velocity variances along the radial direction at $x/D=1.50$ (a) and 3.50 (b), both for $\overline{\lambda}\approx5$. For both cases, $\hat{\lambda}=0.9$ and $St = 0.04$.}
\label{fig:rt_Uvar}
\end{figure}

Finally, to observe the tip-vortex dynamics more directly, radial sweeps of the streamwise-velocity variance at two streamwise locations are presented for $\overline{\lambda}\approx5$ in Fig.\ \ref{fig:rt_Uvar}. While it was not possible to achieve phase-locked measurements of both the low-frequency forcing and tip-vortex shedding, since the turbine was aerodynamically driven, the strength of the tip vortices may still be inferred from the magnitude of the peaks in velocity variance at $r/D\approx\pm0.5$. At $x/D=1.50$ (Fig.\ \ref{fig:rt1-5_Uvar}), variations in tip-vortex strength are evident as a function of phase, with stronger tip vortices observed for $t/T > 0.5$ when the tip-speed ratio is lower (\textit{cf}., Fig.\ \ref{fig:TSR_varyTSR}). Further downstream at $x/D=3.50$ (Fig.\ \ref{fig:rt3-5_Uvar}), signatures of tip-vortex breakdown and the subsequent convergence of shear layers toward the wake centerline are evident. To determine the advection-based phase lag between these two streamwise locations, we use an empirically fitted linear velocity profile of $U/U_\infty \approx 0.18 + 0.2x/3D$ between $1.5\leq x/D \leq 3.5$ and integrate to obtain

\begin{equation}
    \frac{t_2 - t_1}{T} \approx 15 St \ln\left( \frac{x_2/D+2.7}{x_1/D+2.7} \right).
    \label{eqn:phase_offset}
\end{equation}

\noindent Thus, the data at $x/D=3.50$ lag those at $x/D=1.50$ by $\Delta t/T \approx 0.23$. This means that the weaker tip vortices at $0.2\lesssim t/T \lesssim0.5$ in Fig.\ \ref{fig:rt1-5_Uvar} can be traced to the larger dark region in Fig.\ \ref{fig:rt3-5_Uvar} between $0.4\lesssim t/T \lesssim 0.7$. These results imply a connection between tip-vortex breakdown and the time-varying wake structure. Differences in tip-vortex circulation and spacing may affect turbulent entrainment into the wake and thus shift the intermediate-wake region along the streamwise direction, particularly if the changes in the tip-vortex system occur out of phase with the time-varying streamwise-velocity profiles. While some recent studies have investigated tip-vortex breakdown in steady-flow turbine wakes at high Reynolds numbers \citep{pique_understanding_2025,grunwald_effect_2026}, our initial findings suggest that the unsteady control of these dynamics could be explored in future work as a potential active wake-control mechanism for slowly time-varying wakes.


\subsection{Implications for wind-farm modeling and control}\label{sec:results_discussion}

The experimental results demonstrate the importance of wake advection for time-varying wind-turbine wakes and highlight the utility of time-varying turbine control for altering wake structure and evolution. Both of these conclusions hold for slow forcing frequencies ($St=0.04$, with results for $St = 0.02$ and 0.06 shown in App.\ \ref{app:St}) and large-amplitude variations in tip-speed ratio and thrust.

To highlight the practical implications of these slow forcing frequencies, consider a 100-m diameter wind turbine in a 10-$\mathrm{ms^{-1}}$ wind, for which a Strouhal number of 0.04 corresponds to a period of 250 seconds. Typical time scales for both wind-speed variations in the atmospheric boundary layer and wind-turbine control systems are also on the order of minutes \citep{van_der_hoven_power_1957,stull_introduction_1988,starke_dynamic_2024}. In a large wind farm, the effects of these types of disturbances will take time to propagate through the array. Although this study has shown that the advection speed of the disturbances follows the wake velocity, a conservative estimate of the propagation time can be obtained by assuming that the disturbance travels faster, at the mean wind speed. Thus, for a streamwise row of four turbines with a spacing of $10D$, a disturbance at the first turbine will take at least 300 seconds to reach the last turbine. Since this is similar to the forcing period of the disturbance, at any given instant in time each turbine will experience a different phase of the disturbance. If the wind-farm control scheme does not account for these advective effects, and instead operates on spatially or temporally averaged data, the controller state estimate will be inaccurate and thus will not perform optimally in response to the disturbance. This example suggests that the effects of wake advection should be considered even in wind conditions that are nominally statistically stationary, since coherent structures in the atmospheric boundary layer may generate minute-scale oscillations in wind speed that can affect wind-farm operation \citep{van_der_hoven_power_1957,stull_introduction_1988}. Models that seek to parameterize these effects should account for wake advection with velocities related to those in the wake, rather than the free stream. This need not be the centerline velocity; the model and experiments of \citet{wei_wake_2024} suggest that the radially averaged streamwise velocity may be more relevant. However, the fact that radial profiles of the streamwise velocity were collected at only three downstream locations in this work means that we lack the data to determine the most appropriate advection velocity scale.

With regard to wind-farm control, it is worth emphasizing that the two cases explored in \S\ref{sec:results_TSR-varying} ($\overline{\lambda}\approx 4$ and 6) had identical tip-speed-ratio amplitudes and nearly identical thrust coefficients, and both underwent quasi-steady forcing (after accounting for wake advection), and yet they still exhibited different time-varying wake dynamics. The key difference in these cases was the relationship between $C_t$ and $\lambda$ about the mean. The time evolutions of thrust and tip-speed ratio with respect to each other, and not just their mean values, determine the overall time-varying wake structure. This is because different instantaneous combinations of $C_t$ and $\lambda$ produce different wake states, and advection dictates the manner in which these states are distributed in space and time throughout the wake. In this study, the range of accessible wake states was limited by the thrust curve of the turbine. However, variable-pitch turbines can independently manipulate thrust and tip-speed ratio. This extra degree of freedom could allow wake profiles to be arbitrarily tuned as a function of inflow conditions, turbine blade pitch, and generator torque. In essence, this is what is done in active wake-mixing control strategies \citep{meyers_wind_2022}, but these results demonstrate that similar ideas could be implemented at much slower time scales.

%% file: s5_conclusions.tex
\section{Conclusions}\label{sec:conclusions}

In this study, the time-varying wake dynamics of a wind turbine forced by a periodic generator torque are investigated at near utility-scale Reynolds numbers in a pressurized-air wind tunnel. Temporal forcings at low Strouhal numbers are considered to evaluate the effects of large-scale, slowly varying atmospheric disturbances on wake properties. The measurements demonstrate the importance of advection even at slow forcing frequencies; the wake evolution can be modeled as quasi-steady only after the effects of advection by the wake velocity are taken into account. A Lagrangian transformation highlights the nonlinear nature of wake advection, showing that disturbances do not simply advect at the free-stream velocity. Further measurements at different mean tip-speed ratios show that time-resolved control of thrust and tip-speed ratio can be leveraged to manipulate the spatiotemporal evolution of the wake, despite the quasi-steady forcing time scales. These dynamics are directly relevant to wind farms operating in the atmospheric boundary layer, as they occur on similar time scales to those of large-scale coherent structures as well as those of turbine controllers. The results therefore have important implications for steady-flow wake modeling approaches, which may lead to sub-optimal wind-farm performance if advection-related time delays are not included. Furthermore, time-varying controllers may allow for additional degrees of freedom in shaping wind-turbine wakes relative to quasi-steady control paradigms.

Finally, these experiments will serve as a foundation for future measurements involving higher forcing frequencies, which will extend the present findings into the nonlinear regime and investigate the efficacy of active wake-mixing strategies at near utility-scale Reynolds numbers. Interactions between wake meandering modes and imposed turbine forcings will also be studied to determine whether wake instabilities can be amplified or mitigated by active control of the turbine. These explorations will benefit greatly from the quasi-steady baseline considerations and analysis provided by the present work.

%% file: s9_appendices.tex
\FloatBarrier

\appendix
\section{Variations in forcing frequency}\label{app:St}

For the sake of completeness, experimental results in which the forcing frequency and amplitude were varied are included here in appendices.

\begin{figure}
\begin{subfigure}[t]{0.48\textwidth}
\centering
  \includegraphics[width=\textwidth]{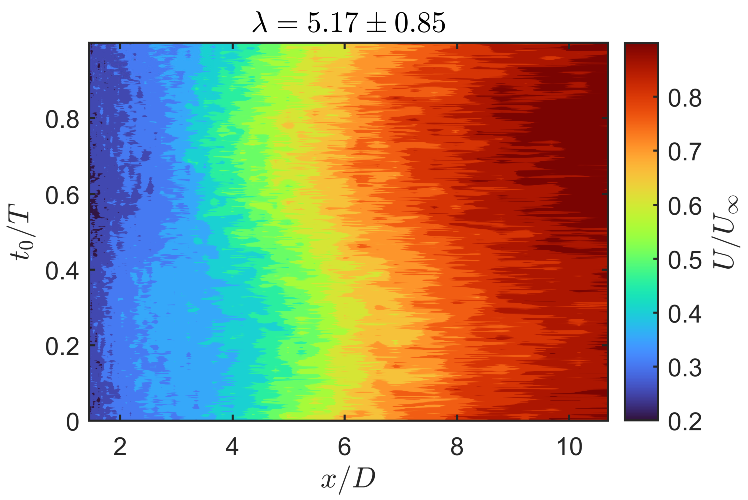}
  \caption{}
\label{fig:app_xt0_U_St1}
\end{subfigure}
\hfill
\begin{subfigure}[t]{0.48\textwidth}
\centering
  \includegraphics[width=\textwidth]{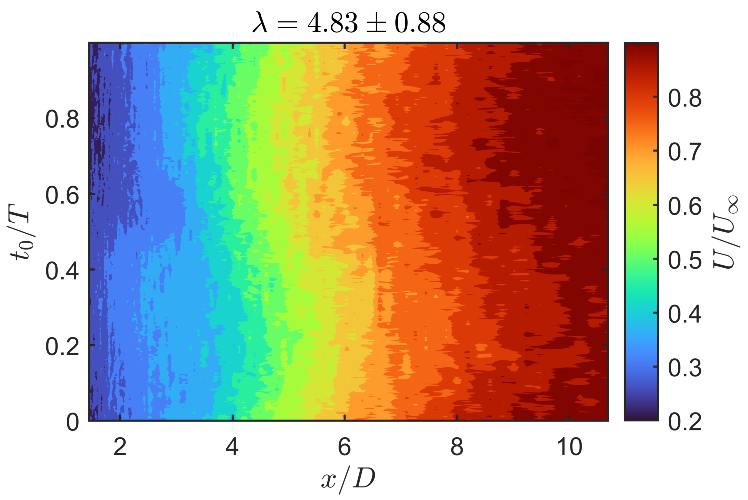}
  \caption{}
\label{fig:app_xt0_U_St3}
\end{subfigure}
\caption{Phase-averaged streamwise velocities along the wake centerline for $\overline{\lambda}\approx5$ and $St = 0.02$ (a) and 0.06 (b), plotted along characteristics defined by an initial time $t_0$.}
\label{fig:app_xt0_U_St}
\end{figure}

Two additional forcing frequencies were tested at $\overline{\lambda}\approx5$ to demonstrate that the dynamics observed in this study are slow enough as to not depend on frequency. These frequencies corresponded to $St=0.02$ and 0.06. For the $St=0.02$ case, data-acquisition limitations meant that 20 periods were included in the phase average for every measurement location. The resulting phase-averaged velocity profiles are shown in Fig.\ \ref{fig:app_xt0_U_St} in the Lagrangian-transformed characteristic space spanned by $t_0/T$ and $x/D$ (see \S\ref{sec:results_time-varying}). The velocity profiles appear very similar to each other, demonstrating that there is little variation in the observed dynamics as a function of forcing frequency. This supports the conclusion that $St=0.04$ is slow enough to count as quasi-steady according to traditional definitions.

\section{Variations in forcing amplitude}\label{app:amp}

\begin{figure}
\begin{subfigure}[t]{0.48\textwidth}
\centering
  \includegraphics[width=\textwidth]{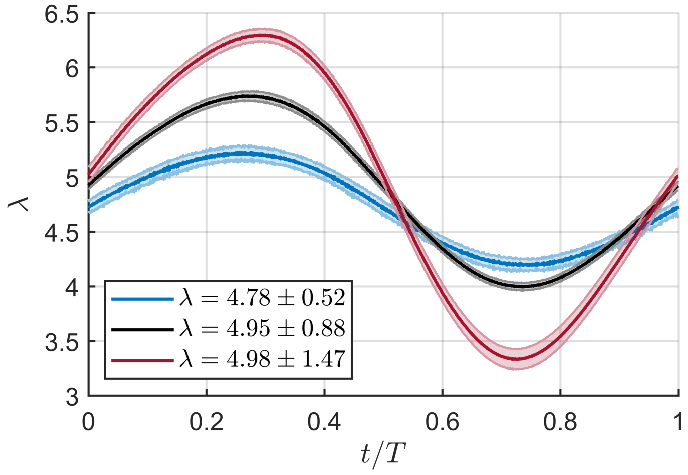}
  \caption{}
\label{fig:app_TSR_varyAmp}
\end{subfigure}
\hfill
\begin{subfigure}[t]{0.48\textwidth}
\centering
  \includegraphics[width=\textwidth]{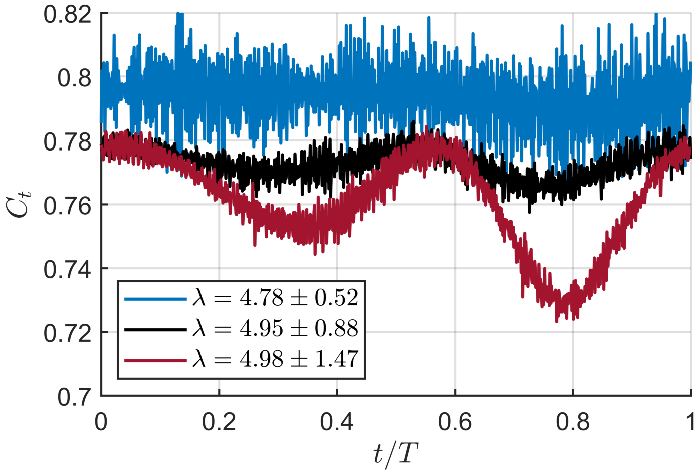}
  \caption{}
\label{fig:app_Ct_varyAmp}
\end{subfigure}
\caption{Phase-averaged tip-speed ratio (a) and thrust coefficient (b) for three tip-speed ratio amplitudes. For all cases, $\overline{\lambda}\approx5$ and $St = 0.04$.}
\label{fig:app_Ct_TSR_varyAmp}
\end{figure}

Two tip-speed ratio amplitudes, in addition to the case of $\hat{\lambda}\approx0.9$, were tested to further investigate the differing effects of thrust and tip-speed ratio on the wake. The resulting tip-speed ratio and thrust-coefficient waveforms are shown in Fig.\ \ref{fig:app_Ct_TSR_varyAmp}. For both cases, $\overline{\lambda}\approx5$ and $St = 0.04$. The low-amplitude case had no significant thrust variation as a function of time relative to measurement error. The high-amplitude case had an effective oscillation period of half the forcing period, due to the decreases in thrust coefficient on both sides of the thrust peak at $\overline{\lambda}\approx5$.

\begin{figure}
\begin{subfigure}[t]{0.48\textwidth}
\centering
  \includegraphics[width=\textwidth]{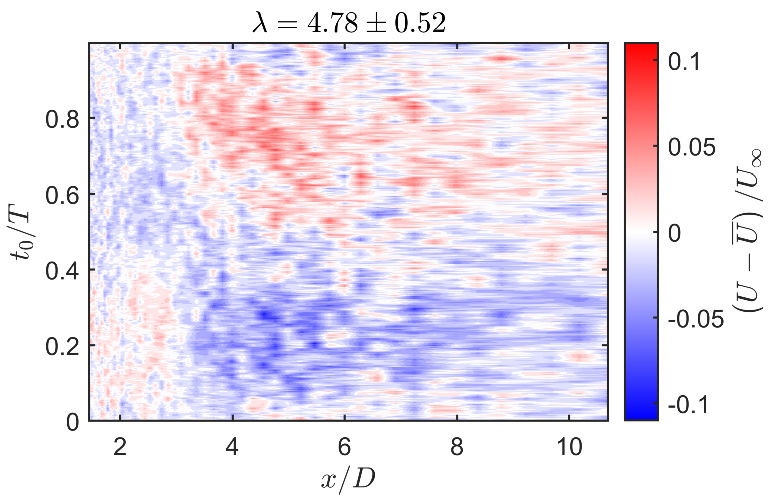}
  \caption{}
\label{fig:app_xt0_Uprime_A1}
\end{subfigure}
\hfill
\begin{subfigure}[t]{0.48\textwidth}
\centering
  \includegraphics[width=\textwidth]{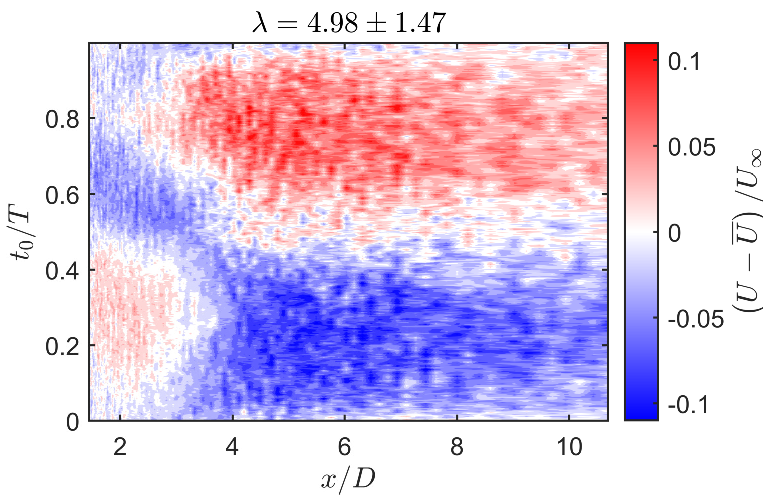}
  \caption{}
\label{fig:app_xt0_Uprime_A3}
\end{subfigure}
\caption{Phase-averaged streamwise velocities along the wake centerline for $\overline{\lambda}\approx5$ and two different tip-speed ratio amplitudes, $\hat{\lambda}\approx0.5$ (a) and $\hat{\lambda}\approx1.5$ (b), plotted along characteristics defined by an initial time $t_0$.}
\label{fig:app_xt0_U_A}
\end{figure}

Fig.\ \ref{fig:app_xt0_U_A} shows the phase-averaged velocity-perturbation waves in the wake, plotted in Lagrangian-transformed characteristic coordinates. The dynamics appear qualitatively similar in all cases, demonstrating that in spite of the relatively large perturbation amplitudes, the wake still behaves in a quasi-steady manner (after accounting for wake advection). For the highest-amplitude case, signatures of the double-peak forcing profile can be seen in the near wake between $2\lesssim x/D \lesssim 4$, where the switch between positive and negative perturbation values occurs twice within each oscillation cycle.

\begin{figure}
\centering
\begin{subfigure}[t]{0.48\textwidth}
  \includegraphics[width=\textwidth]{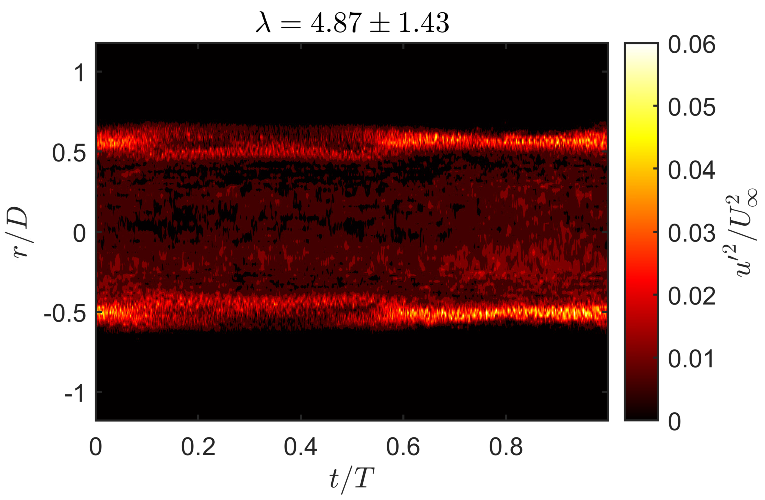}
  \caption{}
\label{fig:app_rt1-5_Uvar_A3}
\end{subfigure}
\begin{subfigure}[t]{0.48\textwidth}
  \includegraphics[width=\textwidth]{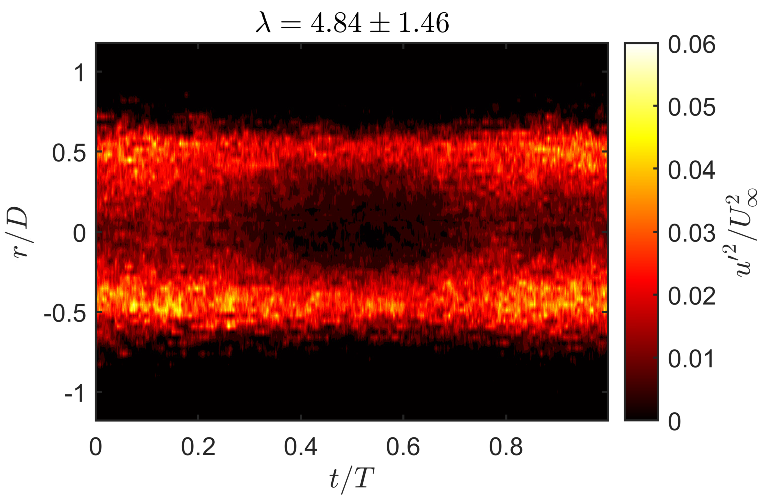}
  \caption{}
\label{fig:app_rt3-5_Uvar_A3}
\end{subfigure}
\caption{Phase-averaged streamwise-velocity variances along the radial direction at $x/D=1.50$ (a) and 3.50 (b), both for $\overline{\lambda}\approx5$ and $\hat{\lambda}=1.4$.}
\label{fig:app_rt_Uvar_A3}
\end{figure}

Radial sweeps of the velocity variance are shown for the highest-amplitude case in \ref{fig:app_rt_Uvar_A3}. The radial profiles show similar behaviors to the middle-amplitude case shown in Fig.\ \ref{fig:rt_Uvar}, including variations in the tip-vortex strength at $x/D=1.50$ and oscillations in the width of the non-turbulent near-wake region at $x/D=3.50$. No strong signatures of a half-period forcing are visible in these data. This suggests that, despite the relatively large thrust amplitude, the shifting tip-vortex breakdown location is the primary driver of near- and intermediate-wake dynamics in this case. These results also show that manipulating the phase and amplitude relationships between $C_t$ and $\lambda$ can lead to widely varying classes of wake-velocity profiles.

